%% file: main.tex
\documentclass{article}
\usepackage[final,main]{neurips_2026}
\usepackage{etoc}
\usepackage{enumitem}
\usepackage[utf8]{inputenc} 
\usepackage[T1]{fontenc}    
\usepackage{hyperref}       
\usepackage{url}            
\usepackage{booktabs}       
\usepackage{amsfonts}       
\usepackage{nicefrac}       
\usepackage{microtype}      
\usepackage[table]{xcolor}  
\usepackage[most]{tcolorbox}
\usepackage{algorithm}
\usepackage{algpseudocode}

\usepackage{graphicx}
\usepackage{amsmath}
\usepackage{amssymb}
\usepackage{hhline}
\usepackage{multirow}
\usepackage{subcaption}
\usepackage{wrapfig}

\usepackage{booktabs}
\usepackage{tabularx}
\usepackage{longtable}
\usepackage{xltabular}
\usepackage{array}
\usepackage{xcolor}
\usepackage{makecell}
\usepackage{caption}
\usepackage{adjustbox}
\usepackage{float}

\definecolor{tableheader}{HTML}{E6F2F5}

\newcommand{\method}{CriticNPU}

\title{Mechanistic Critics for Sample-Efficient NPU Design-Space Exploration}

\author{%
  Yicheng He, Yuqi Xue, Tianyang Wu, Jian Huang, Bin Hu \\
  University of Illinois Urbana-Champaign\\
  \texttt{yh84@illinois.edu} \\
}

\begin{document}
\etocdepthtag.toc{mtchapter}
\maketitle




\begin{abstract}
NPU design-space exploration for LLM inference mainly relies on an expensive simulator that often returns only scalar feedback such as latency and power. 
Such feedback can rank evaluated configurations, but gives little information for deciding why a configuration is slow or which architecture parameter should be changed next. 
This limitation is especially severe because prefill and decode workloads can be limited by different resources, including compute throughput and HBM bandwidth. 
We propose \textbf{\method{}}, a sample-efficient DSE framework built around an online-calibrated mechanistic critic for NPU architecture design. 
The critic uses explicit hardware limits to evaluate NPU configurations: before simulation, it estimates whether a proposed configuration is likely to improve over the current best and filters weak candidates; after simulation, it aggregates operator-level runtimes into prefill/decode bottleneck summaries and predicts the effect of candidate parameter changes. 
We instantiate the critic with a Roofline-style model that derives compute and memory limits from architecture parameters and calibrates per-operator-type effective limits using simulator observations. 
And \method{} combines this critic with a workload extractor and an architecture planner to propose configurations. 
Across six LLM workloads and 16 deployment configurations per model, \method{} achieves best-or-tied-best latency in 74\% of prefill and 87.5\% of decode settings while using roughly half the simulator calls of competitive DSE baselines. 
We conduct extensive ablation studies and detailed analyses to validate the effectiveness of our method from multiple perspectives. 
Our code is available at~\url{https://github.com/bruno686/\method}.
\end{abstract}

\section{Introduction}

Neural processing units (NPUs)~\citep{npu,DianNao,Eyeriss} are specialized accelerators for neural-network workloads and are increasingly important for large language model (LLM) inference~\citep{he2025visplay,llm1,CoEvolution}. 
Given an LLM specification and a deployment setting, NPU design-space exploration (DSE) searches for an NPU configuration that minimizes inference latency under hardware constraints. 
An NPU configuration is defined by architecture parameters such as systolic-array count and dimension, vector-unit count, clock frequency, on-chip memory size, HBM bandwidth, and interconnect bandwidth, while satisfying power and memory constraints~\citep{P3-LLM,TriGen}. 
This search is difficult because architecture parameters interact in coupled and non-smooth ways~\citep{ShadowNPU}. 
For example, adding systolic arrays gives little benefit after the workload becomes memory-bound; increasing HBM bandwidth may not help when inter-chip communication dominates; and lowering frequency can sometimes improve latency if it frees power for more parallel units. 
LLM inference makes the problem harder because prefill and decode often stress different resources, so the best NPU configuration for TTFT may differ from the best configuration for TPOT.

Most DSE methods use simulator to evaluate candidate configurations, but the simulator often returns only scalar feedback such as latencya and power. 
This feedback can rank evaluated configurations, but it gives little help for deciding which architecture parameter should be changed next. 
A slow configuration may be limited by compute throughput, HBM bandwidth, ICI bandwidth, or a mixture of them; scalar latency alone does not reveal which case applies. 
Black-box search methods, including Bayesian optimization, evolutionary search, and greedy search, can work with scalar feedback~\citep{bochip,BOOM-explorer,ConfuciuX}, but they spend simulator calls to learn which regions of the configuration space are useful. 
Bottleneck-aware methods add more structure by analyzing simulator traces or critical paths~\citep{ArchExplorer,zhang2026lumina}. 
They can explain why an evaluated configuration is slow, but the diagnosis is tied to the configuration that was already simulated. 
It does not directly tell the search procedure whether an unevaluated candidate configuration should be simulated, or which architecture parameters should be prioritized in the next update. 
Recent LLM-assisted DSE methods can propose multi-parameter configurations from prompts, search history, and design rules~\citep{ChatDSE,LLM-DSE,LLM-Box}, but deciding whether a proposed configuration is worth simulating still needs a signal from hardware behavior.

This suggests that sample-efficient DSE needs a different form of simulator feedback: one that can reject weak candidates before simulation and turn simulated bottlenecks into concrete constraints on the next parameter update.
We propose \emph{mechanistic critics} for sample-efficient NPU design-space exploration. 
A mechanistic critic evaluates a candidate configuration through explicit hardware limits, rather than only through a scalar simulator score or an LLM-generated estimate. 
For NPU configurations, these limits correspond to compute throughput, HBM bandwidth, and ICI bandwidth. 
The critic is used at two points in the search loop. 
Before simulation, it estimates whether a proposed configuration is likely to improve over the current best configuration, so weak candidates can be skipped before calling the simulator. 
After simulation, it reads the operator-level runtimes, summarizes where prefill and decode latency is spent, and predicts the effect of candidate parameter changes. 
This diagnosis is then used to constrain the next parameter update. 
For example, if the simulated configuration is memory-bound, the next update should not simply add more systolic arrays; it should change parameters that affect memory time. 
Thus, the critic does more than explain evaluated configurations: it also helps choose which configurations to simulate and which parameters to prioritize next.

This paper is not about making a given edge LLM faster. Instead, it addresses a missing first step in edge LLM NPU design: before hardware exploration begins, the system must determine which LLM should become the NPU design target, based jointly on user requirements and hardware constraints.
We instantiate the mechanistic critic in \textbf{\method{}}, a sample-efficient framework for NPU DSE. 
The critic is implemented as an online-calibrated Roofline-style model~\citep{Roofline}: for each NPU configuration, it derives compute and memory limits from architecture parameters and calibrates per-operator-type effective limits using simulator observations. 
\method{} has three components. 
A workload extractor (Computing Demand Agent) maps the LLM specification and deployment setting to an operator-level LLM workload. 
The calibrated mechanistic critic (Online-Calibrated Roofline Model) provides candidate filtering, per-phase bottleneck summaries, and candidate parameter-change predictions.
An LLM-based architecture planner (Architecture Design Agent) proposes the next NPU configuration. 
The planner proposes configurations, while the decision of which candidates to simulate and which architecture parameters to prioritize comes from the calibrated critic. 
Across six LLM workloads and 16 deployment configurations per model, \method{} achieves best-or-tied-best latency in 74\% of prefill and 87.5\% of decode settings while using roughly half the simulator calls of competitive baselines, including Bayesian optimization,
CMA-ES, greedy search, grid search, ChatDSE, LLM-DSE, and LLM-Box. Our contributions are:
\begin{itemize}[leftmargin=10pt]
    \item \textbf{Framing NPU DSE around mechanistic critics.}
    We formulate sample-efficient NPU design-space exploration as a critic-guided search problem, where simulator feedback is used not only to score evaluated configurations, but also to decide which candidate configurations are worth simulating and which architecture parameters should be prioritized next.
    \item \textbf{Calibrating Roofline limits into predictive and diagnostic feedback.}
    We instantiate the critic with a Roofline-style model that derives compute, memory limits from NPU architecture parameters and calibrates per-operator-type effective limits using simulator observations. 
    Before simulation, the critic filters weak candidates; after simulation, it summarizes operator-level runtimes into per-phase bottleneck shares and predicts the effect of candidate parameter changes. 
    \item \textbf{Reducing simulator cost for LLM inference DSE.}
    We build \method{}, a sample-efficient NPU DSE framework that combines workload extraction from an LLM specification and deployment setting, critic-guided candidate filtering, and LLM-based configuration proposal. 
    Across six LLM workloads and 16 deployment configurations per model, \method{} achieves best-or-tied-best latency in 74\% of prefill and 87.5\% of decode settings while using roughly half the simulator calls of competitive DSE baselines.
\end{itemize}

\section{Preliminary}
\paragraph{Problem Formulation.}
\label{sec:problem}
Our goal is to find an NPU configuration with minimal latency $\mathcal{L}$ for a given LLM specification $W$, while satisfying constraints $\mathcal{C}$ including power, HBM bandwidth and capacity, as well as interconnect and datacenter network bandwidth limits.
An NPU configuration $\theta$ is a joint assignment of architecture parameters (systolic array count and size, clock frequency, HBM bandwidth, ICI bandwidth, etc.) drawn from a discrete design space $\mathcal{S}$. Finding the best configuration for LLM inference is a constrained optimization,
\begin{equation}
\min_{\theta \in \mathcal{S}} \;\; \mathcal{L}(\theta; W, D) \quad \text{s.t.} \quad \mathcal{C},
\label{eq:main_opt}
\end{equation}
where $D$ is the deployment setting (e.g., number of parallel chips). Following standard LLM serving practice, the latency $\mathcal{L}$ is decomposed into TTFT (time-to-first-token, the prefill phase) and TPOT (time-per-output-token, the decode phase). A description of various parameters is in Appendix~\ref{parameter_description}.

\paragraph{Roofline Model.}
\label{pre-roofline}
The roofline model~\citep{Roofline} provides a simple yet effective framework for understanding how efficiently hardware can execute a given workload and identifying whether performance is constrained by compute throughput or memory bandwidth.
Specifically, it bounds achievable performance as a function of arithmetic intensity $I = \mathrm{FLOPs}/\mathrm{Bytes}$, peak compute throughput $C$, and memory bandwidth $B$, expressed as $\min(C,\, B \cdot I)$.
The crossover point $I_{\mathrm{ridge}} = C/B$ separates memory-bound operators ($I < I_{\mathrm{ridge}}$) from compute-bound ones ($I > I_{\mathrm{ridge}}$).
When deploying LLM on NPU, this classification identifies the primary hardware bottleneck for each operator, providing a principled basis for adjusting architecture parameters.
For clarity, we provide several illustrative examples in Appendix~\ref{roofline}.


\section{Method}
\begin{figure}[t]
  \centering
  \includegraphics[width=1\textwidth]
  {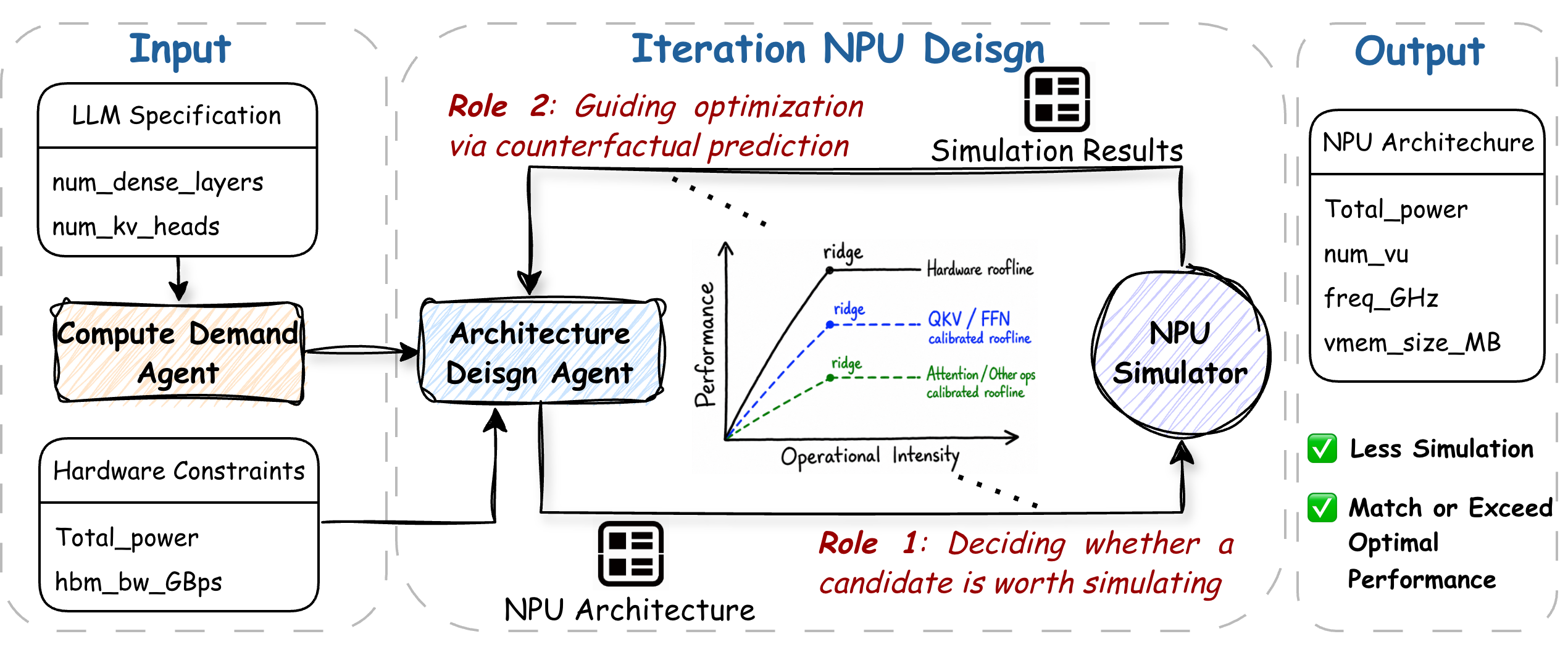}
  \vspace{-0.5cm}
  \caption{Overview of \textbf{\method{}}. A Compute Demand Agent extracts the operator-level workload, an online-calibrated roofline model filters weak configurations and provides bottleneck-guided feedback, and an Architecture Design Agent proposes the next NPU configuration. The shown parameters are illustrative examples.
}
  \label{fig:main}
\end{figure}

\method{} treats NPU design-space exploration as mechanistic critic-guided search over NPU configurations. 
Instead of using simulator feedback only to score evaluated configurations, \method{} maintains an online-calibrated mechanistic critic that works at two points in the search loop: before simulation, it filters weak candidate configurations; after simulation, it summarizes prefill/decode bottlenecks and predicts the effect of candidate parameter changes. 
At the implementation level, \method{} has three components, as shown in Figure~\ref{fig:main}.
\textbf{(1)~Compute Demand Agent}, \underline{implemented as a workload extractor}, maps the LLM specification and deployment setting to an LLM workload. 
\textbf{(2)~Online-Calibrated Roofline Model}, \underline{serving as the calibrated mechanistic critic}, derives hardware limits from each NPU configuration, calibrates per-operator-type effective limits from simulator observations, and provides candidate filtering and bottleneck-guided feedback. 
\textbf{(3)~Architecture Design Agent}, \underline{an architecture planner}, proposes the next NPU configuration using the workload, constraints, search history, bottleneck summaries, and candidate parameter-change predictions. 
We describe these components in the following sections.

\subsection{Compute Demand Agent: LLM-Driven Workload Extractor}
\label{sec:workload}

\paragraph{Why Agent extraction.}
Modern LLM architectures evolve rapidly: dense Transformers, Mixture-of-Experts (DeepSeek-V2/V3), grouped-query attention (Llama-3), Flash-Attention variants, and emerging routing schemes all introduce different operator graphs.
Hand-coding an extractor for each new LLM specification is labor-intensive: it requires re-implementing sharding rules, attention variants, and parallelism-aware tensor reshaping for every model family, and any new release breaks the pipeline until a developer manually updates the code.
A structured workload profile exposes the actual compute and memory demand seen by the NPU, whereas the raw LLM specification mainly describes model dimensions, tensor shapes, and layer counts.
We instead delegate this task to an Agent, the Compute Demand Agent, which reads the LLM specification and produces the corresponding operator-level workload.

\paragraph{How the Agent extracts the workload.}
Given an LLM specification $W$ together with the deployment setting, the Compute Demand Agent produces a per-operator workload $\mathcal{O}$:
\begin{equation}
\mathcal{O} \sim \pi_{\mathrm{ComputeDemandAgent}}(\cdot \mid W, D)
\label{eq:demand_agent}
\end{equation}
where $\pi_{\mathrm{ComputeDemandAgent}}$ denotes an LLM-based policy that generates $\mathcal{O}$ conditioned on $W$ and $D$. Each $\mathcal{O}$ is a set of operators each annotated with its floating-point cost $\mathrm{FLOPs}(o)$, memory traffic $\mathrm{Bytes}(o)$ (weights, activations, and KV cache), execution count $c(o)$ per forward pass, and arithmetic intensity $I(o) = \mathrm{FLOPs}(o)/\mathrm{Bytes}(o)$.
The detailed prompt is provided in Appendix~\ref{prompt_compute_demand}.

\subsection{Online-Calibrated Roofline Model: Mechanistic Critic for Screening and Diagnosis}
\label{sec:wm}
\paragraph{Why a roofline model in the loop.} A roofline model is useful for NPU parameter design because it ties performance to hardware limits that change directly with the configuration. 
Given the compute, memory limits of a candidate configuration, the roofline provides a way to estimate whether latency is likely to improve and to identify which operators limit performance. 
We use this structure to provide performance feedback at two points in the design process. \textit{\textbf{Role 1: Deciding whether a candidate is worth simulating.}} Before invoking the simulator on a proposed configuration, we predict its latency from the roofline. If the prediction clears the current best, the configuration is simulated for ground truth; otherwise it is skipped. This cuts the number of costly simulator calls. \textit{\textbf{Role 2: Guiding optimization via counterfactual prediction.}} After each simulation, we do not hand the raw per-operator runtimes to the Design Agent. The roofline digests them into a structured diagnosis of where time is spent, and predicts how small parameter changes would shift this distribution, so the agent gets directed guidance for its next move, not just raw numbers.

\paragraph{Role 1: Deciding whether a candidate is worth simulating.}
The simulator is the bottleneck of our search, since each call is expensive, and the Design Agent produces many candidates. We therefore want a cheap way to estimate a candidate's latency before paying for a full simulation. The key idea: a recent simulated reference configuration $\theta$ already tells us how long each operator took. A new candidate $\theta'$ that only tweaks the architecture parameters changes these runtimes in a predictable way, as a compute-bound operator gets faster if $\theta'$ has more compute, a memory-bound one gets faster with more bandwidth, and comm-bound ops depend on ICI bandwidth. So predicting $\theta'$ reduces to two things per operator: \emph{what it is bottlenecked on}, and \emph{how much the corresponding hardware resource changes}.

The hardware peaks are $C(\theta) = n_{\mathrm{sa}}\,d_{\mathrm{sa}}^2 \cdot 2 f$ and $B(\theta) = B_{\mathrm{HBM}}$, where $n_{\mathrm{sa}}$ is the number of systolic arrays, $d_{\mathrm{sa}}$ the SA side length, $f$ the clock frequency, and $B_{\mathrm{HBM}}$ the HBM bandwidth---all components of $\theta$. The roofline then gives a single cutoff that separates the two regimes---the \emph{ridge point} $I_{\mathrm{ridge}}(\theta) = C(\theta)/B(\theta)$. For each operator $o$, we compare its arithmetic intensity $I(o)$ to the ridge: $I(o) > I_{\mathrm{ridge}}$ labels it compute-bound, $I(o) < I_{\mathrm{ridge}}$ memory-bound.
Let $T(o;\theta)$ denote the runtime of operator $o$ measured by the simulator at the reference $\theta$. The predicted runtime at $\theta'$ then scales by how the relevant peak changes:
\begin{equation}
T(o;\theta') = T(o;\theta) \cdot
\begin{cases}
C(\theta)/C(\theta') & \text{compute-bound}, \\
B(\theta)/B(\theta') 
& \text{memory-bound}. 
\\
\end{cases}
\label{eq:time_scaling}
\end{equation}
Summing $c(o)\,T(o;\theta')$ over operators and rescaling against the reference TTFT/TPOT gives the predicted latency $\hat{\mathcal{L}}(\theta')$; an analytical power estimate $P(\theta')$ is computed in parallel from $\theta'$.
Let $t$ index the search iteration. Each prediction is coupled to a simple gating rule: simulate $\theta'$ during a short warm-up or when the predicted latency is at most the current best $\mathcal{L}^*$; otherwise skip it.
\begin{equation}
\mathrm{Simulate}(\theta') =
\begin{cases}
1, & t < t_0 \quad\text{(warm-up)}, \\
1, & \hat{\mathcal{L}}(\theta') \le \mathcal{L}^* \quad\text{(predicted no worse than best)}, \\
0, & \text{otherwise (skip)},
\end{cases}
\label{eq:gating_rule}
\end{equation}
where $t_0$ is the warm-up iteration budget. The simulator is thus reserved for candidates the roofline cannot confidently dismiss.

\paragraph{Role 2: Guiding optimization with counterfactual prediction.}
The simulator returns accurate but raw per-operator runtimes. Handing this output directly to the Design Agent is noisy and hard to act on, since the agent cannot easily see where time is going, or whether a given parameter change would help. Role~2 of the roofline is to \emph{digest} each simulation into a compact, actionable signal: (i) where the current configuration is bottlenecked, and (ii) what would happen under small hypothetical edits. \emph{Step 1: diagnose where time goes.} Each operator $o$ is classified using the same test as Role 1: 
otherwise $I(o) > I_{\mathrm{ridge}}(\theta) \Rightarrow$ compute-bound, $I(o) < I_{\mathrm{ridge}}(\theta) \Rightarrow$ memory-bound. The labels are aggregated into per-phase bottleneck shares,
\begin{equation}
\alpha^{\phi}_r(\theta) = \frac{\sum_{o \in \mathcal{O}^\phi_r(\theta)} c(o)\,T(o;\theta)}{\sum_{o \in \mathcal{O}^\phi} c(o)\,T(o;\theta)}, \quad r \in \{\text{comp}, \text{mem}\},\ \phi \in \{\text{prefill}, \text{decode}\},
\label{eq:bottleneck_share}
\end{equation}
so the planner sees, per phase, what fraction of latency sits under each regime. \emph{Step 2: counterfactuals for candidate edits.} For each candidate parameter change the planner is considering, the roofline reuses the scaling rule of Eq.~\ref{eq:time_scaling} to predict its effect, namely which operators speed up, which stay the same, whether the dominant regime shifts, and how TTFT and TPOT would change, all without another simulation. The outputs are written into a short block of text through fixed templates (no LLM inside the roofline) and appended to the Design Agent's prompt. A typical line reads: ``\emph{prefill is currently 60\% compute-bound on FFN-type operators; lowering frequency to free power for two more SAs is predicted to reduce TTFT by $\approx 12\%$ without violating the power budget.}'' The numerical claims here (the share, the speedup, the power check) are deterministic outputs of Eqs.~\ref{eq:bottleneck_share} and \ref{eq:time_scaling}; the LLM only reads them, it does not generate them.

\paragraph{Why the raw peaks are not enough.}
Both roles above classify each operator by where its arithmetic intensity $I(o)$ sits relative to $C(\theta)/B(\theta)$. This implicitly assumes the workload can reach the full compute peak $C$ when compute-bound and the full memory peak $B$ when memory-bound, yet no real workload does. A softmax dependency chain stops the systolic array from filling; LayerNorm has so little math per byte that cache misses dominate; pipeline stalls and imperfect compilation leave headroom the peaks never see. How big this gap is depends on the operator: an FFN GEMM can use most of the systolic array, attention noticeably less, LayerNorm barely any of it. So $C/B$ sits above the true transition for some operator types and can sit below for others, and using it as a single ridge mis-labels operators near the boundary. 
A mis-labelled operator then picks the wrong branch in Eq.~\ref{eq:time_scaling}, scaling by $C$ when it should scale by $B$, or the other way around, which is a predictable source of error that widens as $\theta$ moves away from the reference.

\paragraph{The online-calibrated roofline.}
We fix this by learning, per operator type, how much of each peak the workload actually reaches. We group operators into a small set of types $\tau$ by their role in the transformer block: QKV projection, FFN GEMM, attention kernel, other matmuls, and non-matmul ops (norm, activation, all-reduce), so that operators in the same group use the hardware in similar ways. For each type we maintain two efficiency factors $\eta^{(\tau)}_{\mathrm{comp}}, \eta^{(\tau)}_{\mathrm{mem}} \in (0, 1]$: the fraction of $C$ and $B$ that this type can actually reach. The peaks themselves stay a hardware property; $\eta^{(\tau)}$ only tells us how well this kind of operator uses them. Put another way, $\eta^{(\tau)}$ is a per-type calibration of $C$ and $B$: the peaks do not change, but each operator type effectively sees its own reduced ceilings. The per-type part is what matters, since a single global scale on $C$ and $B$ would only shift one ridge line for all operators, while type-specific $\eta$ gives each operator type its own ridge, so the classifier can label different types differently. Because these factors come from the operator's shape and structure, rather than the chip, they stay roughly the same when we change NPU parameters, and transfer from one configuration to the next.

These factors are learned from simulator feedback. After every simulation, we read each operator's SA time $T_{\mathrm{SA}}(o)$, memory time $T_{\mathrm{mem}}(o)$, FLOP count, and memory traffic, compute the observed per-operator efficiencies
\begin{equation}
\hat{\eta}^{(\tau)}_{\mathrm{comp}}(o) = \frac{\mathrm{FLOPs}(o)/T_{\mathrm{SA}}(o)}{C(\theta)}, \qquad \hat{\eta}^{(\tau)}_{\mathrm{mem}}(o) = \frac{\mathrm{Bytes}(o)/T_{\mathrm{mem}}(o)}{B(\theta)},
\label{eq:eta_observed}
\end{equation}
average them across operators of the same type, and update $\eta^{(\tau)}$ with an exponential moving average $\eta^{(\tau)}_{t+1} = (1-\lambda)\,\eta^{(\tau)}_{t} + \lambda\,\bar{\hat{\eta}}^{(\tau)}_{t}$ (with $\lambda \in (0,1)$), which stays stable against per-operator noise while adapting as new simulations arrive.
The calibration plugs into Roles~1 and 2 through effective ceilings $\tilde{C}^{(\tau)}(\theta) = \eta^{(\tau)}_{\mathrm{comp}}\, C(\theta)$ and $\tilde{B}^{(\tau)}(\theta) = \eta^{(\tau)}_{\mathrm{mem}}\, B(\theta)$, turning the single ridge into a per-type one $\tilde{I}^{(\tau)}_{\mathrm{ridge}}(\theta) = (\eta^{(\tau)}_{\mathrm{comp}}/\eta^{(\tau)}_{\mathrm{mem}})\,C(\theta)/B(\theta)$. The classifier in Roles~1 and 2 now tests each operator against the ridge of its own type, while the scaling rule in Eq.~\ref{eq:time_scaling} and the bottleneck share in Eq.~\ref{eq:bottleneck_share} keep their form, since $\eta^{(\tau)}$ cancels out of the ratios. When $\theta$ changes during search, only $C(\theta)$ and $B(\theta)$ are recomputed; the learned $\eta^{(\tau)}$ carry over.

\subsection{Architecture Design Agent: LLM-Driven Architecture Planner}
\label{sec:planner}

At iteration $t$, the Architecture Design Agent (Prompt template at Appendix~\ref{prompt_arch_design}) observes a compact state summarizing the workload, the current configuration, the counterfactual prediction, the design-space constraints, and the search history,
\begin{equation}
\theta_{t+1} \sim \pi_{\mathrm{ArchDesignAgent}}(\mathcal{O},\, \theta_t,\, \alpha(\theta_t),\, \mathcal{C},\, \mathcal{H}_t),
\label{eq:planner_policy}
\end{equation}
where $\pi_{\mathrm{ArchDesignAgent}}$ is an LLM-based policy that generates the next configuration $\theta_{t+1}$ conditioned on the workload $\mathcal{O}$, the current configuration $\theta_t$, the roofline-guided signal $\alpha(\theta_t)$, hardware constraints $\mathcal{C}$, and the search history $\mathcal{H}_t = \{(\theta_i,\, \mathcal{L}(\theta_i),\, \alpha(\theta_i))\}_{i \le t}$.

\section{Experiments}
To provide a fair and thorough evaluation, we test \method{} on six LLM specifications ranging from 8B to 671B parameters: Llama3-8B, Llama2-13B, Llama3-70B, Llama3.1-405B, DSv2-236B, and DSv3-671B~\citep{touvron2023llama,liu2024deepseek}. These workloads cover both dense Transformers and mixture-of-experts models. We use NeuSIM~\citep{neusim1}, an analytical NPU simulator that reports operator-level latency breakdowns, power, and memory usage for each NPU configuration and workload. We compare against three groups of baselines: black-box optimization methods, including Bayesian optimization (BO)~\citep{bochip} and CMA-ES~\citep{eschip}; search heuristics, including greedy local search~\citep{greedychip} and grid search~\citep{grid_search}; and LLM-assisted DSE methods, including ChatDSE~\citep{ChatDSE}, LLM-DSE~\citep{LLM-DSE}, and LLM-Box~\citep{LLM-Box}. We evaluate all methods across 16 combinations of hardware constraints and deployment settings. Following disaggregated serving, we optimize prefill and decode chips separately: prefill is evaluated by time-to-first-token (TTFT), and decode is evaluated by time-per-output-token (TPOT). TTFT is computed as the latency of processing the input prompt in the prefill stage, while TPOT is computed as the per-token latency of one autoregressive decoding step using the KV cache.

\subsection{Main Results}

\begin{table}[t]
    \centering
    \caption{Win rate and average simulator calls across 16 deployment configurations per model. Small models (8B, 13B) use 8 chips; large models (70B--671B) use 64 chips (DSv2-236B uses 16 chips). Win rate counts configurations where the method achieves the best (or tied-for-best) latency. \textbf{Bold} marks the best value per column (highest Win, lowest Sims); \underline{underline} marks the second-best. Our method (\method{}) is highlighted in gray.}
    \label{tab:main-winrate}
    \small
    \setlength{\tabcolsep}{4pt}
    \resizebox{\linewidth}{!}{
    \begin{tabular}{c lcc|cc|cc|cc|cc|cc}
        \toprule
        \multirow{2}{*}{} & \multirow{2}{*}{Method}
        & \multicolumn{2}{c|}{Llama3-8B}
        & \multicolumn{2}{c|}{Llama2-13B}
        & \multicolumn{2}{c|}{Llama3-70B}
        & \multicolumn{2}{c|}{Llama3.1-405B}
        & \multicolumn{2}{c|}{DSv2-236B}
        & \multicolumn{2}{c}{DSv3-671B}
        \\
        &
        & Win~$\uparrow$ & Sims~$\downarrow$
        & Win~$\uparrow$ & Sims~$\downarrow$
        & Win~$\uparrow$ & Sims~$\downarrow$
        & Win~$\uparrow$ & Sims~$\downarrow$
        & Win~$\uparrow$ & Sims~$\downarrow$
        & Win~$\uparrow$ & Sims~$\downarrow$
        \\
        \midrule

        \multirow{8}{*}{\rotatebox{90}{\textbf{Prefill}}}
        & BO
        & 11/16 & 117
        & 8/16 & 117
        & \underline{12/16} & 117
        & 2/16 & 117
        & 8/16 & 117
        & 10/16 & 117 \\

        & CMA-ES
        & 0/16 & 130
        & 3/16 & 130
        & 3/16 & 130
        & 2/16 & 130
        & 0/16 & 130
        & 0/16 & 130 \\

        & Greedy
        & \textbf{15/16} & 129
        & \textbf{15/16} & 130
        & 10/16 & 130
        & \underline{9/16} & 130
        & \textbf{14/16} & 130
        & \textbf{13/16} & 130 \\

        & Grid
        & 0/16 & 1440
        & 0/16 & 1440
        & 0/16 & 1440
        & 1/16 & 1440
        & 0/16 & 1440
        & 0/16 & 1440 \\

        & ChatDSE
        & 0/16 & $\backslash$
        & 0/16 & $\backslash$
        & 0/16 & $\backslash$
        & 0/16 & $\backslash$
        & 0/16 & $\backslash$
        & 0/16 & $\backslash$ \\

        & LLM-DSE
        & 9/16 & \underline{108}
        & 8/16 & \underline{106}
        & 7/16 & \underline{107}
        & 7/16 & \underline{108}
        & 7/16 & \underline{107}
        & 9/16 & 98 \\

        & LLM-Box
        & 3/16 & 115
        & 5/16 & 117
        & 6/16 & 116
        & 3/16 & 117
        & 5/16 & 117
        & 3/16 & \underline{87} \\

        \rowcolor{gray!15}
        & \textbf{\method{} (Ours)}
        & \underline{14/16} & \textbf{50}
        & \underline{11/16} & \textbf{51}
        & \textbf{13/16} & \textbf{55}
        & \textbf{10/16} & \textbf{54}
        & \underline{11/16} & \textbf{56}
        & \underline{12/16} & \textbf{57} \\

        \midrule
        & Method
        & Win~$\uparrow$ & Sims~$\downarrow$
        & Win~$\uparrow$ & Sims~$\downarrow$
        & Win~$\uparrow$ & Sims~$\downarrow$
        & Win~$\uparrow$ & Sims~$\downarrow$
        & Win~$\uparrow$ & Sims~$\downarrow$
        & Win~$\uparrow$ & Sims~$\downarrow$
        \\
        \midrule

        \multirow{8}{*}{\rotatebox{90}{\textbf{Decode}}}
        & BO
        & \textbf{16/16} & 117
        & \textbf{16/16} & 117
        & \textbf{16/16} & 117
        & \textbf{16/16} & 117
        & \textbf{16/16} & 117
        & \textbf{15/16} & 117 \\

        & CMA-ES
        & \textbf{16/16} & 130
        & \textbf{16/16} & 130
        & \textbf{16/16} & 130
        & \textbf{16/16} & 130
        & 7/16 & 130
        & 10/16 & 130 \\

        & Greedy
        & \textbf{16/16} & 129
        & \textbf{16/16} & 130
        & \textbf{16/16} & 130
        & \textbf{16/16} & 130
        & \textbf{16/16} & 130
        & \underline{14/16} & 130 \\

        & Grid
        & 0/16 & 1440
        & 0/16 & 1440
        & 0/16 & 1440
        & 0/16 & 1440
        & 0/16 & 1440
        & 0/16 & 1440 \\

        & ChatDSE
        & 0/16 & $\backslash$
        & 0/16 & $\backslash$
        & 0/16 & $\backslash$
        & 0/16 & $\backslash$
        & 0/16 & $\backslash$
        & 0/16 & $\backslash$ \\

        & LLM-DSE
        & \textbf{16/16} & \underline{108}
        & \textbf{16/16} & \underline{106}
        & \textbf{16/16} & \underline{107}
        & 14/16 & \underline{108}
        & 12/16 & \underline{107}
        & 8/16 & 98 \\

        & LLM-Box
        & \textbf{16/16} & 115
        & 15/16 & 117
        & \textbf{16/16} & 116
        & 11/16 & 117
        & 9/16 & 117
        & 5/16 & \underline{87} \\

        \rowcolor{gray!15}
        & \textbf{\method{} (Ours)}
        & 15/16 & \textbf{50}
        & \textbf{16/16} & \textbf{51}
        & \textbf{16/16} & \textbf{55}
        & \textbf{16/16} & \textbf{54}
        & 13/16 & \textbf{56}
        & 8/16 & \textbf{57} \\

        \bottomrule
    \end{tabular}}
\end{table}

\begin{table}[t]
    \centering
    \caption{Component ablation of \method{} on Llama3-8B (8 chips, TP=4, PP=2), Llama3-70B (64 chips, TP=8, PP=8), and Llama3.1-405B (64 chips, TP=8, PP=8), all with BS=48, 300W under disaggregated serving. TTFT reported in seconds (prefill chip), TPOT in milliseconds (decode chip); Sims counts simulator calls consumed in each phase.}
    \label{tab:ablation}
    \small
    \setlength{\tabcolsep}{3pt}
    \resizebox{\linewidth}{!}{
    \begin{tabular}{l cc cc cc cc cc cc}
        \toprule
        \multirow{2}{*}{\textbf{Ablation Methods}}
        & \multicolumn{4}{c}{\textbf{Llama3-8B}} & \multicolumn{4}{c}{\textbf{Llama3-70B}} & \multicolumn{4}{c}{\textbf{Llama3.1-405B}} \\
        \cmidrule(lr){2-5} \cmidrule(lr){6-9} \cmidrule(lr){10-13}
        & TTFT & Sims & TPOT & Sims & TTFT & Sims & TPOT & Sims & TTFT & Sims & TPOT & Sims \\
        \midrule
        \textbf{\method{} (Full)}              & 0.4042 & 39 & 9.03 & 15 & 0.4164 & 35 & 15.93 & 13 & 2.0015 & 39 & 71.96 & 13 \\
        \quad w/o Compute Demand Agent        & 0.4042 & 41 & 9.03 & 15 & 0.4248 & 41 & 15.93 & 13 & 2.0285 & 40 & 71.96 & 13 \\
        \quad w/o Roofline model              & 0.4042 & 38 & 9.03 & 48 & 0.4248 & 39 & 15.93 & 38 & 2.0464 & 40 & 71.96 & 41 \\
        \quad Role 1 only (gating, no digest) & 0.4042 & 37 & 9.03 & 15 & 0.4164 & 36 & 15.93 & 13 & 2.0015 & 38 & 71.96 & 15 \\
        \quad Role 2 only (digest, no gating) & 0.4042 & 40 & 9.03 & 42 & 0.4164 & 42 & 15.93 & 35 & 2.0015 & 39 & 71.96 & 44 \\
        \bottomrule
    \end{tabular}}
\end{table}

\textbf{A - \method{} achieves competitive design quality with substantially fewer simulator calls.}
Table~\ref{tab:main-winrate} reports win rate across the 16 deployment settings detailed in Tables~\ref{tab:c1}--\ref{tab:c16} together with the average number of simulator calls per run. \method{} wins 74.0\% of prefill and 87.5\% of decode configurations using only 54 simulator calls on average; every one of its 96 runs stays under 100 calls, a property no other competitive method shares. Classical baselines (BO, CMA-ES, Greedy) reach comparable or higher win rates only by spending 2.2--2.4$\times$ more simulations, while the LLM-based baselines fall behind for opposite reasons: LLM-DSE and LLM-Box query the simulator every iteration but lack a physics prior, so they wander the design space and finalize suboptimally, whereas ChatDSE bypasses simulator feedback and relies on the LLM to estimate latency and constraint satisfaction directly, which leads to configurations that fail to achieve any best-or-tied-best latency in the main-table settings. \method{} avoids both failure modes by using the calibrated roofline to anchor proposals to the currently limiting resource, which is most effective when latency is governed by standard compute/HBM/ICI ceilings. Its main exception is DSv3-671B decode, where BO obtains 15/16 wins while \method{} obtains 8/16. This setting is less well captured by the current roofline because DSv3 decode is not a pure bandwidth/compute scaling problem: MoE routing, activated-expert imbalance, and cross-expert KV-cache/communication effects introduce structure not explicitly modeled by our operator-type calibration. In this regime, BO's larger simulation budget can be more robust by directly sampling these irregular effects; we treat this as a limitation rather than a failure of the overall simulator-efficiency trend. 
As a complement to win rate, we also report mean normalized latency $\overline{L_m / L^*}$ across all 96 cells (where $L^*$ is the best valid latency any method reached on that cell): \method{} achieves $1.012\times$ universal-best on TTFT and $1.008\times$ on TPOT---within $1.2\%$ of optimum on average, comparable to BO ($1.006$) and Greedy ($1.009$) which spend $2.2$--$2.4\times$ more simulator calls, and well below CMA-ES ($1.242$), LLM-Box ($1.198$), LLM-DSE ($1.528$, p90 $2.86$), and ChatDSE ($3.500$). The win-rate gaps among the strongest methods therefore mostly reflect small latency margins rather than large quality differences: BO, Greedy, and \method{} often reach configurations near the same best observed latency. The key difference is simulator cost---\method{} reaches near-best designs with far fewer calls, so its main advantage is sample efficiency.

\begin{figure}[t]
    \centering
    \includegraphics[width=\linewidth]{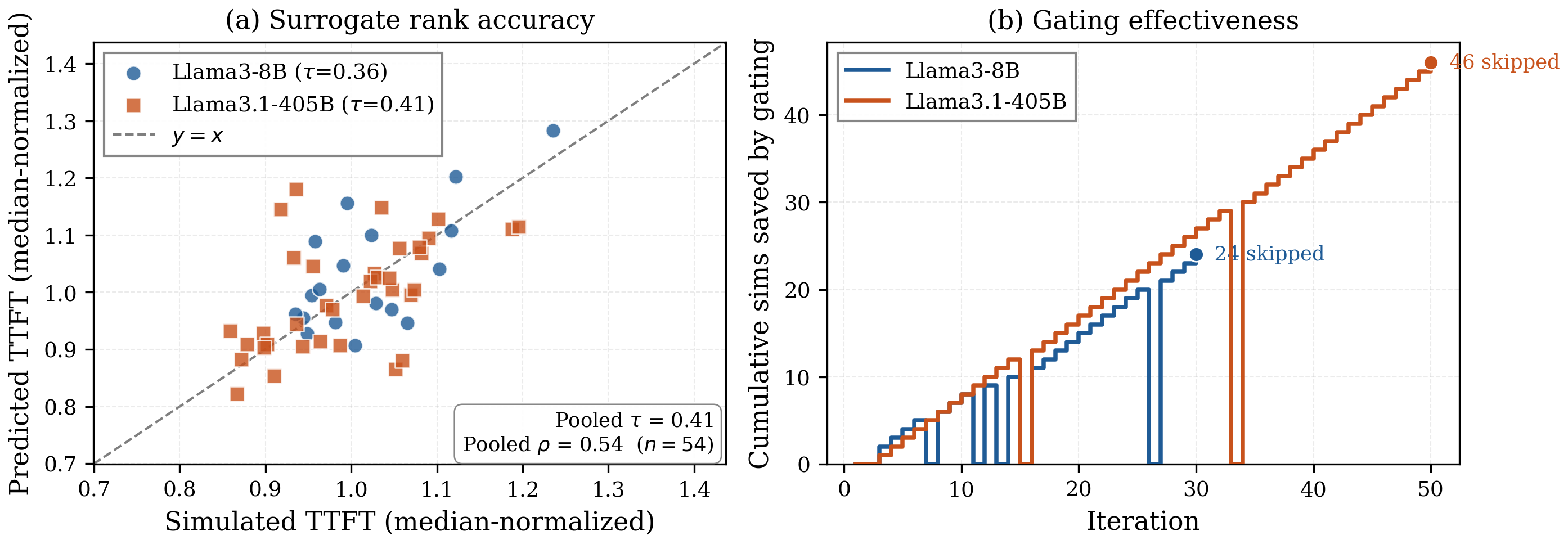}
    \caption{
    Surrogate ranking quality and gating effectiveness.
    \textbf{(a)} Predicted vs.\ simulated TTFT on Llama3-8B and Llama3.1-405B prefill
    (54 proposals total, median-normalized within each workload).
    The calibrated roofline provides a positive rank signal, with Kendall's $\tau=0.36$ and $0.41$
    on the two workloads, respectively, and pooled Kendall's $\tau=0.41$ / Spearman's $\rho=0.54$.
    \textbf{(b)} Cumulative simulator calls skipped by roofline-based gating.
    The gate skips 24 and 46 simulator calls on the two runs, respectively, pruning roughly half of the proposed candidates before simulation.
    }
    \label{fig:surrogate}
\end{figure}

\begin{figure}[t]
    \centering
    \includegraphics[width=1\columnwidth]{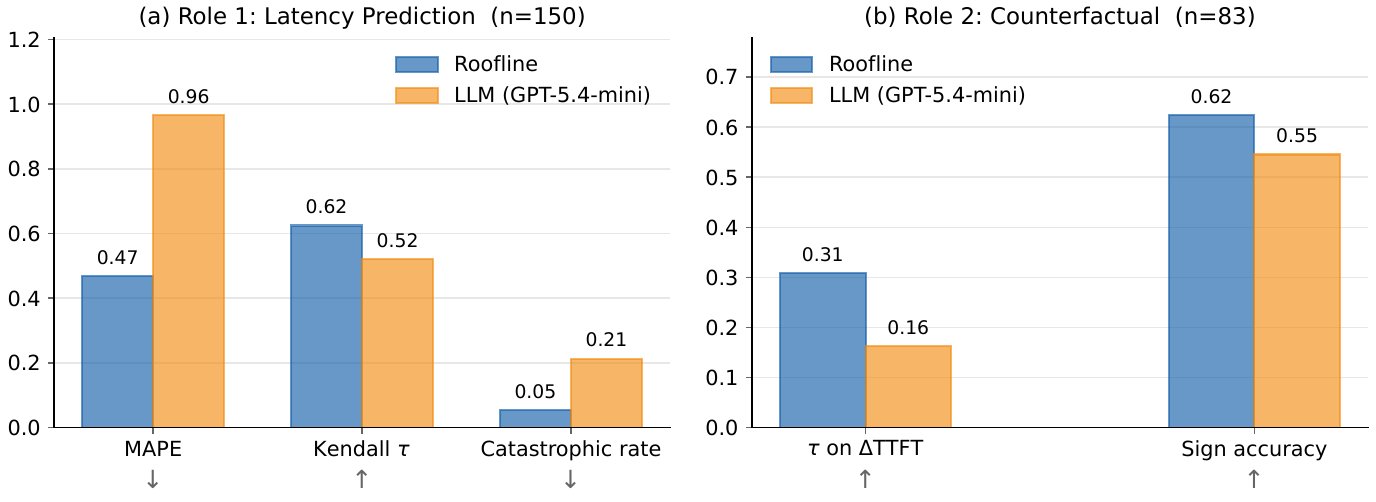}
    \caption{Roofline vs.\ GPT-5.4-mini on latency prediction (Role 1, $n{=}150$) and counterfactual $\Delta$TTFT prediction (Role 2, $n{=}83$). Both methods receive identical inputs: workload, reference config with measured TTFT, and target config. Arrows mark lower/higher-is-better.}
    \label{fig:llm_vs_roofline}
\end{figure}

\subsection{Detailed Analysis}
\paragraph{A - Surrogate Accuracy and Screening Reliability.}
Figure~\ref{fig:surrogate} evaluates what gating actually relies on: the ability of the calibrated roofline to rank configurations, and the sim-budget savings it produces when that ranking is used to screen proposals. Panel (a) plots surrogate-predicted against simulated TTFT on 54 proposals from a 30-iteration run on Llama3-8B prefill and a 50-iteration run on Llama3.1-405B prefill, after normalizing each workload by its median to put both on a common scale. The absolute values are noisy, since the theoretical roofline systematically underestimates latency, but the ordering is positively correlated with the simulator within each workload (Kendall's $\tau=0.36$ on Llama3-8B, $0.41$ on Llama3.1-405B), which is what gating requires. Panel (b) shows the practical consequence: over the course of the two runs, roofline-based gating skipped 24 and 46 simulator calls respectively --- roughly half of every proposal the Design Agent issued. The combined message is that the surrogate does not need to predict absolute latency accurately; it only needs to decide whether a candidate is plausibly worse than the current best, and the adaptive-margin gating rule (Eq.~\ref{eq:gating_rule}) uses the moderate but positive ranking signal to prune a large fraction of simulator calls without degrading final design quality (cf.\ Table~\ref{tab:main-winrate}).

\paragraph{B - Can an LLM Replace the Roofline?}
Figure~\ref{fig:llm_vs_roofline} asks whether GPT-5.4-mini, the same model \method{} uses as the Design Agent, can replace the roofline on the two prediction tasks it performs in the loop. Given identical inputs (workload, reference config with measured TTFT, target config) on 150 Role-1 pairs and 83 Role-2 pairs drawn from the main-table runs, the roofline wins on every metric: 47\% vs.\ 96\% MAPE and 5\% vs.\ 21\% catastrophic errors on absolute prediction, and $\tau=0.62$ vs.\ $0.52$ on Role 1 ranking and $\tau=0.31$ vs.\ $0.16$ on Role 2 $\Delta$TTFT ranking. The LLM's absolute MAPE looks only twice as bad because it hedges by echoing the reference TTFT, since this strategy carries almost no information about which proposed edit is actually an improvement, which is exactly what gating and counterfactual reasoning need. We therefore keep the LLM for high-level design planning and let the roofline do the quantitative surrogate work.

\paragraph{C - Effect of Online Calibration.}
\begin{wrapfigure}{r}{0.46\columnwidth}
    \vspace{-0.8em}
    \centering
    \includegraphics[width=0.46\columnwidth]{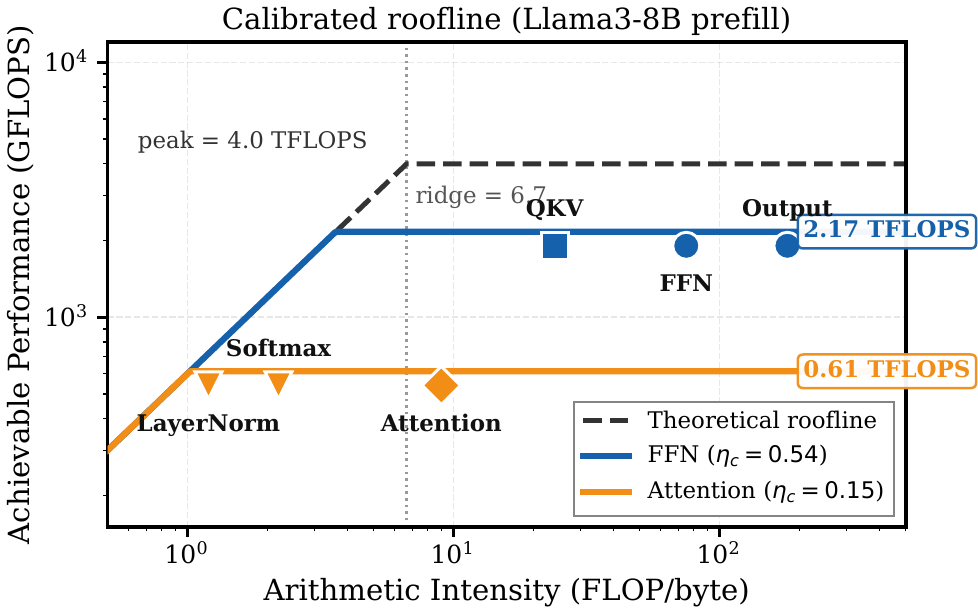}
    \caption{Calibrated roofline for Llama3-8B prefill. Per-type $\eta$ pulls the FFN ceiling to 54\% and attention to 15\% of peak, shifting the effective ridge.}
    \label{fig:roofline}
    \vspace{-0.8em}
\end{wrapfigure}
Figure~\ref{fig:roofline} shows the effect of online calibration on Llama3-8B prefill: the per-type efficiency factors converge to $\eta_{\mathrm{FFN}}{=}0.54$ and $\eta_{\mathrm{attention}}{=}0.15$, placing the two operator classes under two different effective ceilings rather than one shared theoretical peak. The gap between the two ceilings is not uniform: FFN operators give up about half their peak to cache misses and pipeline stalls, while attention operators forfeit nearly six times more because the softmax dependency serializes work that the matrix-multiply ceiling assumes runs in parallel. The resulting ridge shifts flip the regime label of boundary operators such as layer-normalization and softmax, giving the Design Agent quantitatively correct thresholds rather than a single optimistic peak that would mistakenly tag memory-bound ops as compute-bound. This is what lets the gating rule in Eq.~\ref{eq:gating_rule} prune proposals without losing the good configurations in Table~\ref{tab:main-winrate}, and it is the same calibration signal Role 2 feeds back into the prompt, so the Design Agent's next move is grounded in what the hardware actually delivers instead of what it could in principle reach.

\paragraph{D - Additional Analyses.}
Due to page limitations, we defer eight additional analyses to the appendix.
Appendix~\ref{app:conv} shows that \method{} converges to near-best designs with fewer simulator calls.
Appendix~\ref{app:hparam} shows stable final TTFT/TPOT across roofline hyperparameters, including gating, EMA, and stopping settings.
Appendix~\ref{app:llmgen} shows that the method generalizes across different agent backbones.
Appendix~\ref{app:per_deployment} reports per-deployment results and confirms the aggregate trends across all constraint and deployment settings.
Appendix~\ref{Workload_Extraction_Fidelity} validates the CDA operator profiles used by the roofline against NeuSIM's native generator.
Appendix~\ref{Equal_Budget_comparative} shows that \method{} reaches near-optimal valid designs faster than baselines under the same simulation budget.
Appendix~\ref{app:false_reject} shows that roofline gating does not discard meaningfully better candidates.
Appendix~\ref{app:sim_robustness} shows that \method{} remains stable under moderate simulator-noise perturbations.

\section{Conclusion and Limitations}
\label{sec:limitations}

We presented \method{}, a sample-efficient NPU DSE framework built around an online-calibrated mechanistic critic. 
The critic turns simulator feedback into practical search guidance: before simulation, it filters weak candidate configurations; after simulation, it summarizes prefill/decode bottlenecks and predicts the effect of candidate parameter changes. 
Across diverse LLM workloads and deployment settings, \method{} reaches near-best latency while using substantially fewer simulator evaluations than competitive baselines.

\textbf{Limitations.}
Our evaluation is conducted entirely in simulation, without physical implementation or tape-out validation. 
The results therefore depend on the fidelity of the NPU simulator; online calibration can reduce mismatch within the simulator, but cannot correct systematic simulator errors. 
The current critic mainly models compute throughput, HBM bandwidth, and inter-chip communication, and may be less accurate for unmodeled effects such as MoE routing imbalance, cross-expert communication, or compiler scheduling artifacts.

\section*{Acknowlegement}
This work was partially supported by the Hybrid Cloud and AI program at the IBM-Illinois Discovery Accelerator Institute (IIDAI), and NSF under the grant CAREER CNS-2144796.

\medskip

\bibliographystyle{unsrtnat} 
\bibliography{ref}

\newpage

\appendix
\onecolumn
\begin{center}
	\LARGE \bf {Appendix of ``NPU Architecture Design''}
\end{center}

\etocdepthtag.toc{mtappendix}
\etocsettagdepth{mtchapter}{none}
\etocsettagdepth{mtappendix}{subsection}
\tableofcontents

\newpage

\section{Roofline Model: A Concrete Example}
\label{roofline}
\begin{figure}[h]
  \centering
  \includegraphics[width=0.7\textwidth]{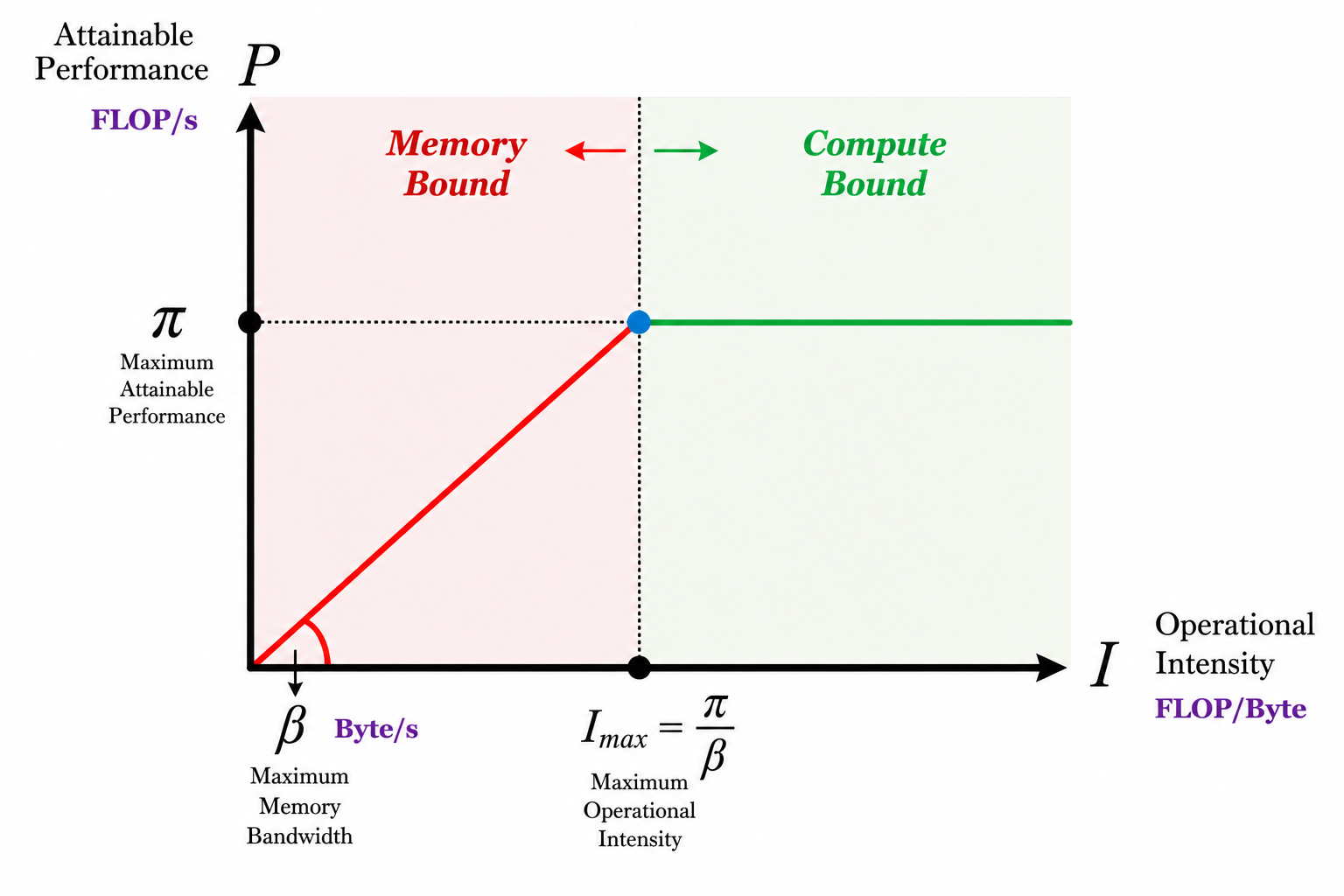}
  \vspace{-0.5cm}
  \caption{Roofline model.}
  \label{fig:roofline_model}
\end{figure}

\label{app:roofline_example}
\newcommand{\ttt}[1]{\texttt{#1}}
To make the roofline model more concrete, we consider a simple example.

\paragraph{Hardware.}
Assume a hardware platform with peak compute throughput $C(\theta) = 10$ TFLOP/s, HBM bandwidth $B(\theta) = 2$ TB/s, and inter-chip communication bandwidth $B_{\mathrm{ici}}(\theta) = 0.5$ TB/s.
The compute-memory ridge point is then
$$
I_{\mathrm{ridge}}(\theta) = \frac{C(\theta)}{B(\theta)} = \frac{10}{2} = 5.
$$

\paragraph{Three operators.}
Now consider three operators.
Operator A has arithmetic intensity $I(o_A)=2$ and no inter-chip transfer.
Operator B has arithmetic intensity $I(o_B)=20$ and no inter-chip transfer.
Operator C has arithmetic intensity $I(o_C)=10$, with $1$ TB of inter-chip transfer and $2$ TFLOPs of computation.

\paragraph{Performance prediction.}
Using the roofline model, operator A has predicted throughput
\[
P(o_A) = I(o_A)\,B(\theta) = 2 \times 2 = 4 \text{ TFLOP/s}.
\]
Since $I(o_A) < I_{\mathrm{ridge}}(\theta)$, operator A is \textbf{memory-bound}.

Operator B has predicted throughput
\[
P(o_B) = C(\theta) = 10 \text{ TFLOP/s}.
\]
Since $I(o_B) > I_{\mathrm{ridge}}(\theta)$, operator B is \textbf{compute-bound}.

\paragraph{Interpretation.}
This example shows that low-intensity operators are limited by HBM bandwidth, high-intensity operators are limited by compute throughput, and operators with large inter-chip transfer can be limited by communication bandwidth even when their arithmetic intensity is high.

\paragraph{Important caveat.}
The values above represent ideal upper bounds.
In practice, the achieved performance is often lower due to factors such as limited data reuse, kernel inefficiencies, imperfect parallelism, and communication overhead.

\section{Parameters Descriptions}
\label{parameter_description}

\subsection{LLM Specification}

\begingroup
\small
\renewcommand{\arraystretch}{1.15}

\begin{xltabular}{\textwidth}{
    >{\ttfamily\raggedright\arraybackslash}p{0.32\textwidth}
    >{\raggedright\arraybackslash}X
    >{\raggedleft\arraybackslash}p{0.18\textwidth}
}
\caption{LLM specification parameters of DeepSeek-V2-236B.}
\label{tab:llm_specification} \\

\toprule
\rowcolor{tableheader}
\textbf{Name} &
\textbf{Description} &
\textbf{Value} \\
\midrule
\endfirsthead

\toprule
\rowcolor{tableheader}
\textbf{Name} &
\textbf{Description} &
\textbf{Value} \\
\midrule
\endhead

model\_name & Model name & deepseekv2-236b \\
model\_type & Model type & llm \\
global\_batch\_size & Global batch size & 1 \\
input\_seqlen & Input sequence length & 4096 \\
output\_seqlen & Output sequence length & 512 \\
decode\_width & Decode width & 1 \\
d\_model & Hidden dimension & 5120 \\
num\_heads & Number of attention heads & 128 \\
d\_head & Dimension of each attention head & 128 \\
d\_ff & Dense FFN intermediate dimension & 12288 \\
moe\_d\_ff & MoE expert intermediate dimension & 1536 \\
num\_layers & Number of transformer layers & 60 \\
num\_dense\_layers & Number of dense layers & 1 \\
ffn\_type & FFN type & deepseek\_moe \\
num\_kv\_heads & Number of KV heads & 128 \\
router\_type & Router type & topk \\
num\_shared\_experts & Number of shared experts & 2 \\
num\_routed\_experts & Number of routed experts & 160 \\
num\_activated\_routed\_experts \\ \_per\_token & Activated routed experts per token & 6 \\
num\_limited\_groups & Number of limited expert groups & 3 \\
q\_lora\_rank & LoRA rank of query projection & 1536 \\
kv\_lora\_rank & LoRA rank of key/value projection & 512 \\
qk\_nope\_head\_dim & Non-RoPE QK head dimension & 128 \\
qk\_rope\_head\_dim & RoPE QK head dimension & 64 \\
v\_head\_dim & Value head dimension & 128 \\
use\_flash\_attention & Whether flash attention is enabled & false \\
data\_parallelism\_degree & Data parallelism degree & 1 \\
tensor\_parallelism\_degree & Tensor parallelism degree & 1 \\
pipeline\_parallelism\_degree & Pipeline parallelism degree & 1 \\
expert\_parallelism\_degree & Expert parallelism degree & 1 \\
num\_data\_parallel\_axes & Number of data-parallel axes & 0 \\
num\_tensor\_parallel\_axes & Number of tensor-parallel axes & 0 \\
num\_pipeline\_parallel\_axes & Number of pipeline-parallel axes & 0 \\
num\_expert\_parallel\_axes & Number of expert-parallel axes & 0 \\
data\_parallel\_degree\_dcn & Data parallel degree across DCN & 1 \\
tensor\_parallel\_degree\_dcn & Tensor parallel degree across DCN & 1 \\
pipeline\_parallel\_degree\_dcn & Pipeline parallel degree across DCN & 1 \\
expert\_parallel\_degree\_dcn & Expert parallel degree across DCN & 1 \\
microbatch\_size\_dcn & Microbatch size for DCN execution & 1 \\
microbatch\_size\_ici & Microbatch size for ICI execution & 1 \\
num\_chips & Number of chips & 1 \\
output\_file\_path & Output CSV file path & v2.csv \\

\bottomrule
\end{xltabular}
\endgroup

\subsection{LLM Workload}

\begin{table}[H]
\centering
\caption{Example workload-level operator profile.}
\label{tab:llm_workload}
\renewcommand{\arraystretch}{1.25}
\begin{adjustbox}{max width=\textwidth}
\begin{tabularx}{\textwidth}{
    >{\raggedright\arraybackslash}p{0.14\textwidth}
    >{\raggedright\arraybackslash}p{0.10\textwidth}
    >{\raggedright\arraybackslash}p{0.15\textwidth}
    >{\raggedright\arraybackslash}X
    >{\raggedright\arraybackslash}X
}
\toprule
\rowcolor{tableheader}
\textbf{Description} &
\textbf{Count} &
\textbf{FLOP Count} &
\textbf{Input Tensor Shapes} &
\textbf{Output Tensor Shapes} \\
\midrule
FFN-up & 32 & $4.81 \times 10^{11}$ & $[16, 4096, 4096]$ & $[16, 4096, 14336]$ \\
\bottomrule
\end{tabularx}
\end{adjustbox}
\end{table}

\subsection{Simulator Outcome}

\begin{table}[H]
\centering
\caption{Simulator output fields and their meanings.}
\label{tab:simulator_outcome}
\renewcommand{\arraystretch}{1.2}
\begin{adjustbox}{max width=\textwidth}
\begin{tabularx}{\textwidth}{
    >{\raggedright\arraybackslash}p{0.18\textwidth}
    >{\ttfamily\raggedright\arraybackslash}p{0.32\textwidth}
    >{\raggedright\arraybackslash}X
}
\toprule
\rowcolor{tableheader}
\textbf{Category} &
\textbf{Field} &
\textbf{Meaning} \\
\midrule

Latency / Throughput & TTFT\_sec & Time to first token, i.e., prefill-phase latency in seconds. \\
Latency / Throughput & TPOT\_ms\_request & Time per output token, i.e., per-token decode latency in milliseconds. \\
Latency / Throughput & throughput\_tokens\_per  \\ \_sec\_request & Throughput in tokens per second per request. \\

\midrule

Memory & mem\_footprint\_GB\_prefill & Memory usage during the prefill phase. \\
Memory & mem\_footprint\_GB\_decode & Memory usage during the decode phase. \\
Memory & hbm\_size\_GB & HBM capacity. \\
Memory & out\_of\_memory & Whether the workload exceeds the HBM capacity. \\

\midrule

Power & avg\_power\_W\_prefill & Average power during prefill. \\
Power & avg\_power\_W\_decode & Average power during decode. \\
Power & max\_power\_W\_prefill & Maximum single-operator instantaneous power during prefill. \\
Power & max\_power\_W\_decode & Maximum single-operator instantaneous power during decode. \\
Power & total\_power\_peak\_W & Analytical chip peak power derived from configuration rather than simulation. \\

\midrule

Config Info & chip\_overrides & The chip configuration evaluated by the simulator. \\
Config Info & output\_dir & Directory where raw outputs are saved. \\

\midrule

Raw Operator Data & prefill\_stats & Aggregate prefill statistics. \\
Raw Operator Data & decode\_stats & Aggregate decode statistics. \\
Raw Operator Data & prefill\_ops & Per-operator breakdown for prefill, including execution time, systolic-array time, memory time, interconnect time, FLOP count, memory traffic, bounded resource, FLOP utilization, and HBM bandwidth utilization. \\
Raw Operator Data & decode\_ops & Same as \texttt{prefill\_ops}, but for decode. \\

\bottomrule
\end{tabularx}
\end{adjustbox}
\end{table}

\subsection{NPU Configuration and Constraints}

\begin{table}[H]
\centering
\caption{NPU architecture parameters and hardware constraints.}
\label{tab:npu_configuration}
\renewcommand{\arraystretch}{1.22}
\begin{adjustbox}{max width=\textwidth}
\begin{tabularx}{\textwidth}{
    >{\ttfamily\raggedright\arraybackslash}p{0.27\textwidth}
    >{\raggedright\arraybackslash}X
    >{\raggedright\arraybackslash}p{0.24\textwidth}
}
\toprule
\rowcolor{tableheader}
\textbf{Name} &
\textbf{Explanation} &
\textbf{Range / Constraint} \\
\midrule

num\_sa & Number of systolic arrays & $2$--$16$ \\
num\_vu & Number of vector units & $2$--$12$ \\
sa\_dim & Systolic array dimension, i.e., $N \times N$ & $64 / 128 / 256$ \\
freq\_GHz & Operating frequency & $0.5$--$2.0$ GHz \\
vmem\_size\_MB & Vector memory size & $32$--$512$ MB \\

\midrule
hbm\_bw\_GBps & HBM bandwidth budget & Specified constraint \\
hbm\_size\_GB & HBM capacity budget & Specified constraint \\
ici\_bw\_GBps & Inter-chip interconnect bandwidth budget & Specified constraint \\
dcn\_bw\_GBps & Data center network bandwidth budget & Specified constraint \\

\midrule

static\_power\_W\_per\_sa & Static power per systolic array & $p_{\mathrm{static}}$ \\
static\_power\_W\_per\_vu & Static power per vector unit & $p_{\mathrm{static}}$ \\
static\_power\_vmem\_W & Static power of vector memory & $p_{\mathrm{static}}$ \\
dynamic\_power\_W\_per\_SA & Dynamic power per systolic array & $p_{\mathrm{dynamic}}$ \\
dynamic\_power\_W\_per\_VU & Dynamic power per vector unit & $p_{\mathrm{dynamic}}$ \\
dynamic\_power\_vmem\_W & Dynamic power of vector memory & $p_{\mathrm{dynamic}}$ \\

\midrule

total\_chip\_power\_W & Total chip power constraint & Specified constraint \\

\bottomrule
\end{tabularx}
\end{adjustbox}
\end{table}

\section{Related Work}

\subsection{Architecture-level Design Space Exploration.}
Architecture-level design space exploration (DSE) directly searches over hardware architectures or microarchitectural configurations, where each candidate corresponds to a different processor, GPU, NPU, or accelerator design. For CPU microarchitecture, BOOM-Explorer~\citep{BOOM-explorer} searches RISC-V BOOM microarchitectural parameters at the RTL level and improves sample efficiency by incorporating microarchitecture domain knowledge into black-box optimization. ArchExplorer~\citep{ArchExplorer} introduces bottleneck-aware microarchitecture exploration by constructing microexecution critical paths and reallocating hardware resources according to the identified bottlenecks. ChatDSE~\citep{ChatDSE} explores BOOM microarchitectures with GPT-based zero-shot reasoning and pseudo labels to reduce the need for expensive labeled samples. Beyond CPUs, LUMINA~\citep{zhang2026lumina} performs LLM-guided GPU architecture exploration for LLM workloads by extracting architectural knowledge from simulator code and sensitivity studies. Several works target DNN or AI accelerator architectures. ConfuciuX~\citep{ConfuciuX} formulates DNN accelerator design as hardware resource assignment and uses reinforcement learning with genetic search to allocate compute and buffer resources under dataflow and platform constraints. AIRCHITECT~V2~\citep{AIRCHITECTv2} treats accelerator DSE as a learning-based recommendation problem, using transformer-based unified representations to recommend accelerator configurations for given workloads and mappings. DiffAxE~\citep{DiffAxE} casts accelerator design as a generative problem and uses a diffusion model to generate accelerator configurations conditioned on target performance or energy-delay characteristics. ArchAgent~\citep{ArchAgent} is also related to architecture-level exploration, but focuses on discovering new architectural mechanisms, such as cache replacement policies, rather than only tuning predefined parameters. \method{} belongs to this architecture-level category and is most closely related to bottleneck-aware methods such as ArchExplorer and LUMINA. However, unlike CPU critical-path analysis or GPU rule extraction, \method{} targets NPU microarchitectural design for LLM inference and uses an online-calibrated roofline model~\citep{Roofline} to provide operator-level compute, memory, and communication bottleneck signals. This enables candidate filtering before simulation and structured design guidance after simulation.

\subsection{Implementation- and Mapping-level Design Space Exploration.}
A second line of work performs DSE above the architecture level, where the hardware substrate, accelerator template, or compilation flow is largely fixed, and the search focuses on how a workload is implemented, compiled, mapped, or deployed. Recently, LLM for engineering~\citep{guo2024controlagent,guokevian,guo2025toward} has become popular.
LLM-DSE~\citep{LLM-DSE} uses a multi-agent LLM framework to optimize HLS directive parameters, such as pipelining, tiling, and parallelization choices, for domain-specific accelerators. iDSE~\citep{iDSE} also targets HLS design optimization, using LLM-guided initialization, pruning, and quality-aware refinement to navigate directive configurations. These HLS-level methods can affect the generated circuit quality, but their design variables are directives mediated through a synthesis toolchain rather than explicit architecture parameters such as compute arrays, SRAM capacity, bandwidth, or interconnect resources. At the system and mapping level, MLDSE~\citep{MLDSE} provides an infrastructure for exploring multi-level hardware organizations and spatiotemporal mappings across hierarchical accelerators, while LLM-Box~\citep{LLM-Box} guides black-box optimization for mapping LLMs onto specialized accelerator systems through choices such as parallelism, collective communication, KV-cache sharding, and interconnect usage. These works optimize implementation or deployment strategies on a given or partially specified substrate. In contrast, \method{} directly searches NPU microarchitectural parameters and uses calibrated performance modeling to reason about how changing hardware resources affects operator-level bottlenecks.

\subsection{Performance Modeling and Bottleneck Analysis}

A complementary line of work develops structured models to capture the relationship between workload characteristics and hardware performance.
Analytical models such as the roofline~\citep{Roofline} provide a physics-grounded abstraction of compute and memory limits, while simulator-based methods perform critical-path or bottleneck analysis to attribute runtime to specific hardware resources~\citep{ArchExplorer}.
Recent work further leverages LLMs to extract design rules from simulator traces or source code~\citep{zhang2026lumina}.
These approaches expose fine-grained performance structure that is not available from scalar metrics alone, and have been used to interpret simulator outputs or derive design heuristics.
However, such models are typically applied after simulation, or used as static predictors that do not adapt to observed execution behavior.
Consequently, they provide diagnostic information about existing configurations but are not used to guide exploration before simulator evaluation. Our online-calibrated roofline can also be viewed as a physics-informed surrogate specialized to architecture search; unlike statistical surrogate-assisted optimization that requires many initial samples, the roofline becomes useful after as few as 3 calibration rounds due to its strong physics prior.
This separation between prediction and optimization limits their effectiveness in reducing the number of simulator calls required for design space exploration.

\section{System Prompt Template}
\label{app:prompt}

\definecolor{outline}{HTML}{3D8CA3}
\definecolor{myblue}{HTML}{C1DBE4}
\definecolor{back}{HTML}{E6F2F5}

\subsection{Prompt for Computing Demand Agent}
\label{prompt_compute_demand}
\begin{tcolorbox}[
  enhanced,
  breakable,
  colback=back,
  colbacktitle=myblue,
  colframe=outline,
  coltitle=black,
  boxrule=1.5pt,
  title=Compute Demand Agent,
]

\small

\textbf{Role Overview}

You are a computing demand analyst for neural accelerator simulation.

Given an LLM model config, parallelism setup, and reference operator-description template, you MUST derive ALL numeric values (Count, FLOP Count, tensor shapes) by applying the formulas below step by step. Do NOT guess or approximate - compute exactly.

\vspace{0.5em}
\textbf{Output Format}

Call the \ttt{compute\_demand} tool with TWO keys - \ttt{derived\_values} first, then \ttt{operators}:

\ttt{
\{
  "derived\_values": \{
    "B": <number>,
    "d\_model\_shard": <number>,
    "Nh": <number>,
    "d\_ff\_chip": <number>,
    "layers\_per\_stage": <number>,
    "decode\_tokens": <number>,
    "gathered\_dim": <number or null>,
    "q\_proj\_b\_out": <number or null>,
    "wo\_input\_dim": <number or null>,
    "moe\_ff\_chip": <number or null>,
    "shared\_ff\_chip": <number or null>,
    "kv\_cache\_seq": <number or null>
  \},
  "operators": [...]
\}
}

You MUST compute \ttt{derived\_values} FIRST from the config, then use ONLY those values in all operators.

\vspace{0.5em}
\textbf{Shape Format Rules}

\begin{itemize}
\item Input shapes: \ttt{DT\_BFLOAT16:[d1,d2,...]}
\item Output shapes: \ttt{[DT\_BFLOAT16:(d1,d2,...)]}
\item Multiple inputs are comma-separated
\item Default dtype: \ttt{DT\_BFLOAT16}
\item Special \ttt{DT\_FLOAT32} ops:
  \begin{itemize}
    \item RMSNorm / LayerNorm (q\_norm, kv\_norm)
    \item Softmax (gate-scores\_softmax)
    \item Rotary embedding (apply\_rotary\_emb)
  \end{itemize}
\item Batch: \ttt{B = global\_batch\_size / (dp * pp)}
\item Sequence: Prefill L = input\_seqlen, Decode L = 1
\end{itemize}

\vspace{0.5em}
\textbf{Dimension Formulas}

\ttt{
tp = tensor\_parallelism\_degree \\
pp = pipeline\_parallelism\_degree \\
ep = expert\_parallelism\_degree \\
dp = data\_parallelism\_degree
}

\begin{itemize}
\item \ttt{B = global\_batch\_size / (dp * pp)}
\item \ttt{d\_model\_shard = d\_model / tp}
\item \ttt{Nh = num\_heads / tp}
\item \ttt{Nkv = num\_kv\_heads / tp}
\item \ttt{d\_ff\_chip = d\_ff / tp}
\end{itemize}

\textbf{DeepSeek MLA (ffn\_type = deepseek\_moe)}

\begin{itemize}
\item \ttt{gathered\_dim = d\_model * tp}
\item \ttt{q\_proj\_b\_out = Nh * (qk\_nope\_head\_dim + qk\_rope\_head\_dim)}
\item \ttt{wo\_input\_dim = Nh * v\_head\_dim}
\item \ttt{moe\_ff\_chip = moe\_d\_ff * ep / tp}
\item \ttt{shared\_ff\_chip = (d\_ff / num\_shared\_experts) / tp}
\item \ttt{kv\_cache\_seq = input\_seqlen + output\_seqlen}
\end{itemize}

\vspace{0.5em}
\textbf{Count Formulas}

\begin{itemize}
\item \ttt{layers\_per\_stage = CEIL(num\_layers / pp)}
\end{itemize}

Standard LLM:
\begin{itemize}
\item PREFILL: ICI = 1
\item DECODE: ICI = decode\_tokens
\end{itemize}

DeepSeek MoE:
\begin{itemize}
\item Attention uses \ttt{num\_layers}
\item FFN uses \ttt{layers\_per\_stage}
\end{itemize}

\vspace{0.5em}
\textbf{FLOP Formulas}

\begin{itemize}
\item Linear: \ttt{2 * B * L * M * N}
\item Batched einsum: \ttt{2 * B * L * Nh * D1 * S}
\item Per-head einsum: \ttt{2 * B * L * Nh * D1 * D2}
\item RMSNorm: \ttt{6 * product(input)}
\item LayerNorm: \ttt{8 * product(input)}
\item Softmax: \ttt{4 * product(input)}
\item FlashAttention: \ttt{4 * B * Nh * L * L * d\_head}
\end{itemize}

\vspace{0.5em}
\textbf{Tensor Shape Patterns}

\textbf{Standard LLM}

\begin{itemize}
\item AllGather: \ttt{d\_model\_shard -> d\_model}
\item FFN uses \ttt{d\_ff\_chip = d\_ff / tp}
\item Q/K/V weights are 3D
\end{itemize}

\textbf{DeepSeek MLA}

\begin{itemize}
\item AllGatherMHA: \ttt{d\_model -> gathered\_dim}
\item Sequence gather along time dimension
\item Residual uses \ttt{d\_model}
\end{itemize}

\vspace{0.5em}
\textbf{Common Mistakes}

\begin{enumerate}
\item Use CEIL, not floor
\item Attention uses num\_layers
\item MLA uses gathered\_dim
\item Nh = num\_heads / tp
\item moe\_ff\_chip = moe\_d\_ff * ep / tp
\item RMSNorm = factor 6
\item Softmax = factor 4
\item Use commas between tensors
\end{enumerate}

\vspace{0.5em}
\textbf{Final Instructions}

\begin{itemize}
\item Compute \ttt{derived\_values} first
\item Use ONLY derived values
\item Keep operator descriptions EXACT
\item Verify:
  \begin{itemize}
  \item \ttt{Nh = num\_heads / tp}
  \item \ttt{gathered\_dim = d\_model * tp}
  \end{itemize}
\item Call \ttt{compute\_demand}
\end{itemize}

\end{tcolorbox}

\subsection{Prompt for the Architecture Design Agent}
\label{prompt_arch_design}

\definecolor{outline}{HTML}{fa7e23}
\definecolor{myblue}{HTML}{FECBAC}
\definecolor{back}{HTML}{FFECE0}

\begin{tcolorbox}[
  enhanced,
  breakable,
  colback=back,
  colbacktitle=myblue,
  colframe=outline,
  coltitle=black,
  boxrule=1.5pt,
  title=Architecture Design Agent,
]

\small

\textbf{Role Overview}

You are an expert NPU (Neural Processing Unit) chip architect.

Your task is to design an NPU chip architecture for a given model, while keeping total power within budget.

\vspace{0.5em}
\textbf{Optimization Goal}

\ttt{\{focus\}}

\vspace{0.5em}
\textbf{Upper-Bound Constraints}

These parameters represent technology limits. You may set them, but they MUST NOT exceed maximums.

\ttt{\{upper\_bounds\_json\}}

You should typically use max values for best performance, but lowering them may reduce power or area.

\vspace{0.5em}
\textbf{Power Budget (Hard Constraint)}

\begin{itemize}
\item Peak power must stay under \ttt{\{power\_budget\_W\}} W
\item \ttt{total\_power\_peak\_W = static + dynamic}
\item If exceeded → configuration is INVALID
\item Reduce power via:
  \begin{itemize}
  \item fewer compute units
  \item lower per-unit power
  \end{itemize}
\end{itemize}

\vspace{0.5em}
\textbf{Baseline Configuration}

\ttt{\{base\_chip\_config\_json\}}

\vspace{0.5em}
\textbf{Tunable Parameters (Full Search)}

\begin{itemize}
\item \ttt{num\_sa}: 2–16
\item \ttt{num\_vu}: 2–12 (ports = num\_vu)
\item \ttt{sa\_dim}: 64 / 128 / 256
\item \ttt{freq\_GHz}: 0.5–2.0
\item \ttt{vmem\_size\_MB}: 32–512
\item \ttt{TDP\_W}: 100–500
\end{itemize}

Power parameters:

\ttt{\{power\_params\_json\}}

\vspace{0.5em}
\textbf{Power Formula}

\ttt{
total\_power\_peak\_W = static + dynamic
}

\begin{itemize}
\item static:
\ttt{(static\_power\_W\_per\_sa * num\_sa) + (static\_power\_W\_per\_vu * num\_vu) + fixed}
\item dynamic:
\ttt{(dynamic\_power\_W\_per\_SA * num\_sa) + (dynamic\_power\_W\_per\_VU * num\_vu) + fixed}
\end{itemize}

Approximation:

\ttt{
total approx fixed + SA\_term + VU\_term + vmem
}

\vspace{0.5em}
\textbf{Key Trade-offs}

\begin{itemize}
\item More SAs → more compute, more power
\item Higher freq → more speed, quadratic power increase
\item Larger SA dim → better large matmuls
\item Decode is often memory-bound BUT:
  \begin{itemize}
  \item too few SAs wastes HBM
  \item usually need num\_sa = 8–12
  \end{itemize}
\item Prefill → compute-bound → increase SAs
\item Larger VMEM → less HBM traffic
\item IMPORTANT: baseline is NOT always optimal
\end{itemize}

\vspace{0.5em}
\textbf{Roofline Model}

\begin{itemize}
\item Compute ceiling:
\ttt{num\_sa * sa\_dim\textasciicircum 2 * 2 * freq\_GHz}
\item BW ceiling:
\ttt{hbm\_bw\_GBps}
\item Ridge point:
\ttt{compute / BW}
\end{itemize}

Interpretation:

\begin{itemize}
\item intensity > ridge → compute-bound
\item intensity < ridge → memory-bound
\end{itemize}

Design strategy:

\begin{itemize}
\item compute-bound → increase compute
\item memory-bound → increase BW or reuse
\item watch ridge shifts
\item bottlenecks cascade
\end{itemize}

\vspace{0.5em}
\textbf{Exploration Rules}

\begin{itemize}
\item NEVER repeat previous configs
\item Explore drastically different regions
\end{itemize}

\vspace{0.5em}
\textbf{Important Notes}

\begin{itemize}
\item \ttt{num\_vu\_ports = num\_vu}
\item Power scales with unit count
\item Check memory vs HBM capacity
\end{itemize}

\vspace{0.5em}
\textbf{Final Instruction}

Call \ttt{propose\_npu\_config} with your proposed configuration.

\end{tcolorbox}

\section{Experiments}
\label{app:experiments}

The extended experiments in this section cover the three questions most likely to follow from the main-text detailed analysis: whether the claimed simulator-budget savings show up as a faster \emph{trajectory} in addition to a lower final point (D, Sec.~\ref{app:conv}), whether the results are robust to the roofline's hyperparameters (E, Sec.~\ref{app:hparam}), and whether \method{}'s advantage transfers across different base LLMs (F, Sec.~\ref{app:llmgen}). Each subsection leads with a headline figure and follows with a per-workload table, so a reviewer can skim each experiment in a single page.

\subsection{Convergence Trajectories}
\label{app:conv}
\begin{figure}[t]
    \centering
    \includegraphics[width=0.98\columnwidth]{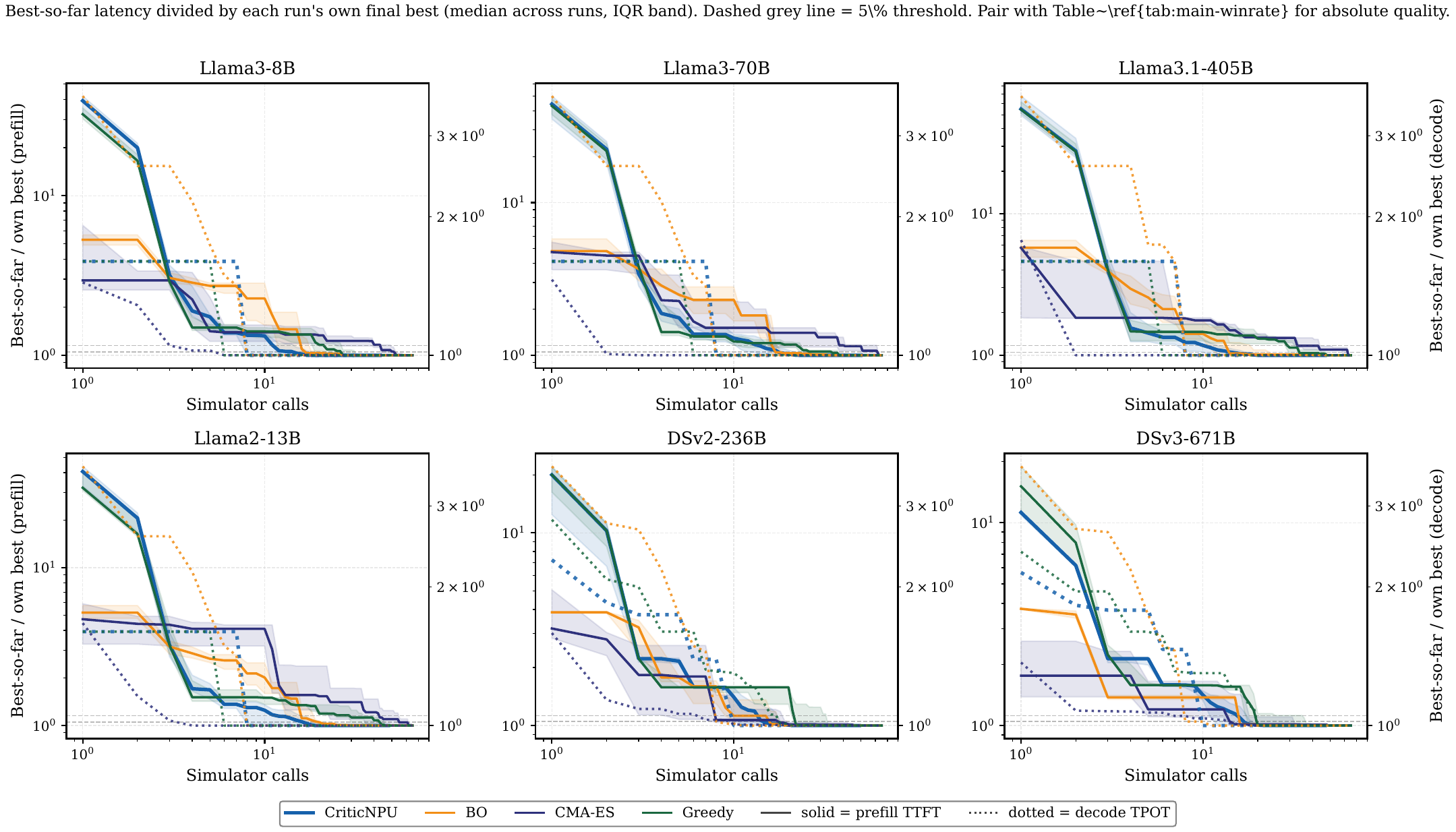}
    \caption{Best-so-far latency vs.\ simulator calls, normalized by each run's own final best (median across that method's main-table runs, IQR band). Solid = prefill TTFT, dotted = decode TPOT. Dashed grey line marks the 5\% threshold. Per-run normalization measures convergence speed only; pair with main table for absolute quality.}
    \label{fig:convergence}
\end{figure}

\begin{figure}[t]
    \centering
    \includegraphics[width=0.96\columnwidth]{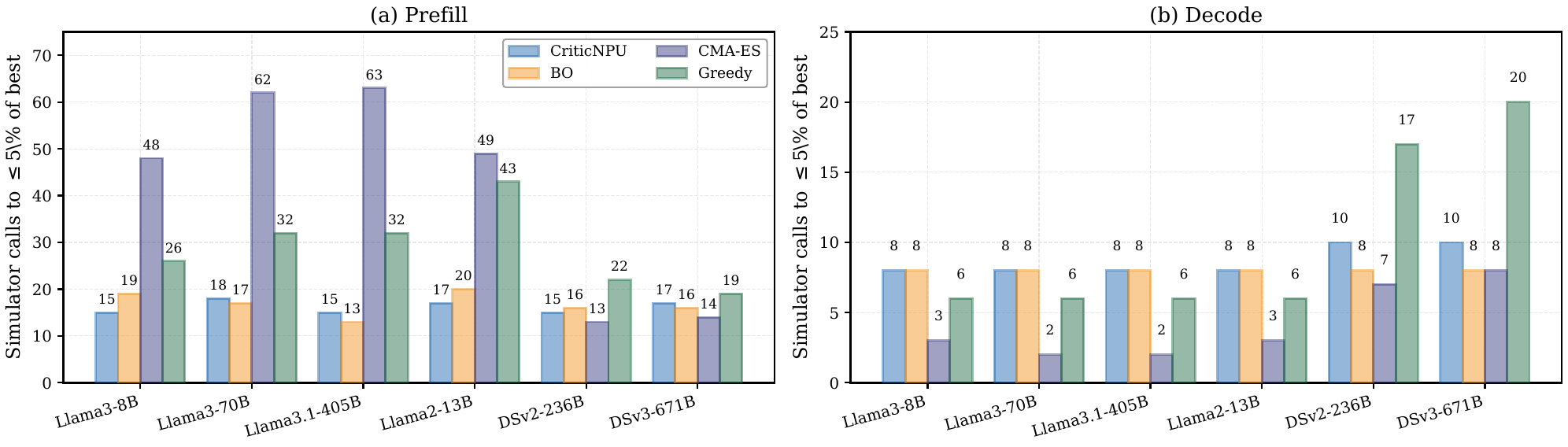}
    \caption{At-a-glance summary of Table~\ref{tab:sims_to_threshold}: simulator calls required to come within 5\% of each method's own final best TTFT (a, prefill) and TPOT (b, decode) on each workload. CMA-ES's fast convergence on decode is a local-minimum artefact --- its decode win rate in Table~\ref{tab:main-winrate} is 0--3 of 16.}
    \label{fig:sims_to_5}
\end{figure}

Figure~\ref{fig:convergence} plots best-so-far latency against cumulative simulator calls across all six workloads, normalized so each run's final best equals 1.0 (lower = closer to its own optimum). Each curve is the median over that method's main-table runs (agent: 41--62 runs per workload-phase; baselines: 2--16 runs depending on coverage), with IQR bands across runs. Table~\ref{tab:sims_to_threshold} reports the median number of simulator calls required to come within $\{20\%, 10\%, 5\%\}$ of the method's own final best. Figure~\ref{fig:sims_to_5} summarizes the $\leq$5\% column as a grouped bar chart. \emph{This measures convergence speed, not absolute quality}; final TTFT / TPOT are already covered by Table~\ref{tab:main-winrate}. A method that converges fast here but has a low win rate there (e.g., CMA-ES on decode) is simply settling into a local minimum quickly. \method{}'s prefill trajectories close to within 5\% of their own best in 15--18 simulator calls across all workloads. Compared with BO and Greedy, \method{} is consistently competitive or faster in prefill, while CMA-ES shows workload-dependent behavior: it converges slowly on several Llama workloads but reaches its own-best threshold quickly on DSv2 and DSv3. In decode, several baselines also reach their own final best within a few calls, which reinforces that this threshold analysis should be interpreted together with the absolute win rates in Table~\ref{tab:main-winrate}. Taken together, the convergence curves and main win-rate results show that \method{} reaches high-quality configurations with substantially fewer simulator calls.

\begin{table}[h]
    \centering
    \caption{Median simulator calls required for each method to come within $\{20\%, 10\%, 5\%\}$ of its own final best TTFT (prefill) / TPOT (decode). ``--'' means the method never reaches the threshold. Lower is better; best per column in bold. Per-run normalization; absolute quality in Table~\ref{tab:main-winrate}.}
    \label{tab:sims_to_threshold}
    \small
    \setlength{\tabcolsep}{3pt}
    \resizebox{\textwidth}{!}{
    \begin{tabular}{l l ccc ccc ccc ccc ccc ccc}
        \toprule
        & & \multicolumn{3}{c}{\textbf{Llama3-8B}} & \multicolumn{3}{c}{\textbf{Llama3-70B}} & \multicolumn{3}{c}{\textbf{Llama3.1-405B}} & \multicolumn{3}{c}{\textbf{Llama2-13B}} & \multicolumn{3}{c}{\textbf{DSv2-236B}} & \multicolumn{3}{c}{\textbf{DSv3-671B}} \\
        \cmidrule(lr){3-5}\cmidrule(lr){6-8}\cmidrule(lr){9-11}\cmidrule(lr){12-14}\cmidrule(lr){15-17}\cmidrule(lr){18-20}
        \textbf{Phase} & \textbf{Method} & $\leq$20\% & $\leq$10\% & $\leq$5\% & $\leq$20\% & $\leq$10\% & $\leq$5\% & $\leq$20\% & $\leq$10\% & $\leq$5\% & $\leq$20\% & $\leq$10\% & $\leq$5\% & $\leq$20\% & $\leq$10\% & $\leq$5\% & $\leq$20\% & $\leq$10\% & $\leq$5\% \\
        \midrule
        \multirow{4}{*}{\rotatebox{90}{\textbf{Prefill}}}
          & NPUAgent & \textbf{11} & \textbf{12} & \textbf{15} & \textbf{13} & \textbf{16} & 18 & \textbf{10} & \textbf{12} & 15 & \textbf{11} & \textbf{15} & \textbf{17} & 13 & 14 & 15 & \textbf{14} & 16 & 17 \\
          & BO & 14 & 14 & 19 & 16 & 17 & \textbf{17} & 12 & 13 & \textbf{13} & 15 & 17 & 20 & 10 & 16 & 16 & 16 & 16 & 16 \\
          & CMA-ES & 26 & 48 & 48 & 38 & 51 & 62 & 33 & 44 & 63 & 34 & 45 & 49 & \textbf{5} & \textbf{6} & \textbf{13} & \textbf{14} & \textbf{14} & \textbf{14} \\
          & Greedy & 20 & 23 & 26 & 14 & 23 & 32 & 31 & 32 & 32 & 20 & 40 & 43 & 22 & 22 & 22 & 18 & 19 & 19 \\
        \midrule
        \multirow{4}{*}{\rotatebox{90}{\textbf{Decode}}}
          & NPUAgent & 8 & 8 & 8 & 8 & 8 & 8 & 8 & 8 & 8 & 8 & 8 & 8 & 9 & 10 & 10 & 10 & 10 & 10 \\
          & BO & 8 & 8 & 8 & 8 & 8 & 8 & 8 & 8 & 8 & 8 & 8 & 8 & 8 & 8 & 8 & 8 & 8 & \textbf{8} \\
          & CMA-ES & \textbf{3} & \textbf{3} & \textbf{3} & \textbf{2} & \textbf{2} & \textbf{2} & \textbf{2} & \textbf{2} & \textbf{2} & \textbf{2} & \textbf{3} & \textbf{3} & \textbf{2} & \textbf{3} & \textbf{7} & \textbf{2} & \textbf{2} & \textbf{8} \\
          & Greedy & 6 & 6 & 6 & 6 & 6 & 6 & 6 & 6 & 6 & 6 & 6 & 6 & 13 & 15 & 17 & 15 & 18 & 20 \\
        \bottomrule
    \end{tabular}}
\end{table}

\subsection{Hyperparameter Sensitivity}
\label{app:hparam}
\begin{figure}[t]
    \centering
    \includegraphics[width=0.98\columnwidth]{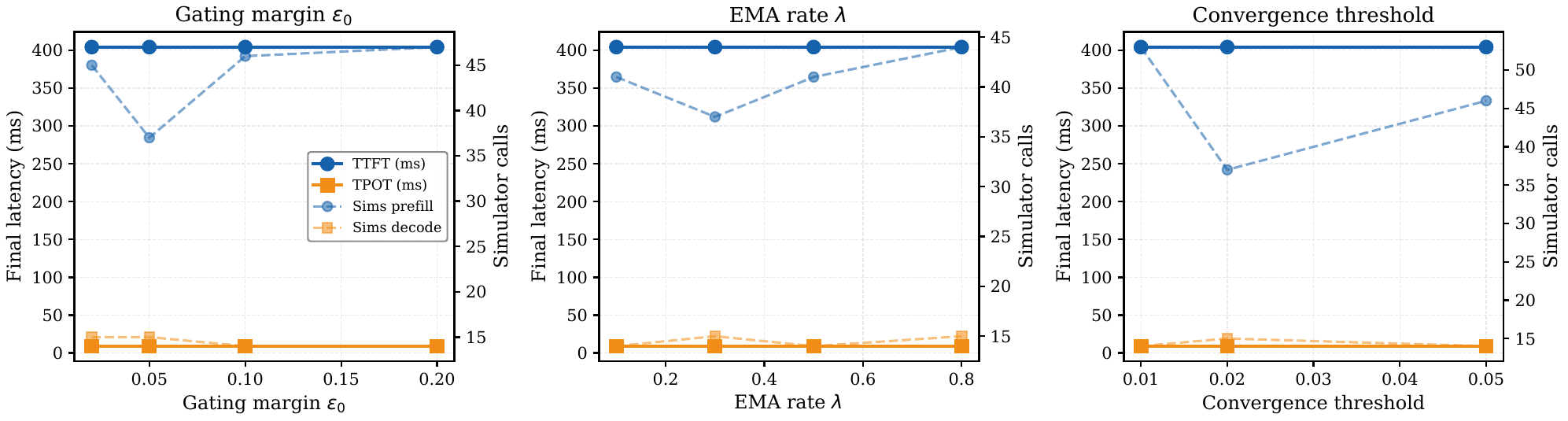}
    \caption{Hyperparameter sensitivity curves on Llama3-8B disaggregated serving (BS=48, 300\,W). Each panel sweeps one knob with the other two held at the default. Solid markers = final TTFT / TPOT (left axis, linear); dashed = simulator-call count per phase (right axis). Final quality is flat across the full range of all three knobs.}
    \label{fig:hparam_curves}
\end{figure}

Three knobs in the roofline module are non-obvious and deserve a robustness check: the gating-margin base $\epsilon_0$ (how aggressively Role 1 skips proposals), the EMA rate $\lambda$ (how quickly learned $\eta^{(\tau)}$ absorb new observations), and the outer convergence threshold $\tau_{\mathrm{conv}}$ (the relative-change stopping criterion). Figure~\ref{fig:hparam_curves} sweeps each knob on Llama3-8B (BS=48, 300W, disaggregated) with everything else held at the default, plotting final TTFT (prefill), final TPOT (decode) and the per-phase simulator-call count on twin axes. Table~\ref{tab:hparam_sensitivity} lists the underlying numbers, with the default setting in bold. Final TTFT / TPOT are identical across the entire sweep ($\Delta=0\%$ in every cell); simulator-call counts vary moderately with the threshold ($\tau_{\mathrm{conv}}$=0.01$\to$53 prefill sims; $\tau_{\mathrm{conv}}$=0.05$\to$46), and barely with $\epsilon_0$ or $\lambda$. The method therefore neither relies on a tuned sweet spot nor degrades under off-default choices of the three knobs.

\begin{table}[t]
    \centering
    \caption{Hyperparameter sensitivity on Llama3-8B disaggregated serving (BS=48, 300\,W). One run per setting; default in bold. $\Delta$TTFT is relative change vs.\ the default.}
    \label{tab:hparam_sensitivity}
    \small
    \setlength{\tabcolsep}{4pt}
    \begin{tabular}{l l cc c cc c c}
        \toprule
        & & \multicolumn{2}{c}{\textbf{Prefill}} & & \multicolumn{2}{c}{\textbf{Decode}} & & \\
        \cmidrule(lr){3-4}\cmidrule(lr){6-7}
        \textbf{Knob} & \textbf{Value}
          & \textbf{TTFT (ms)} & \textbf{Sims} &
          & \textbf{TPOT (ms)} & \textbf{Sims} &
          & \textbf{$\Delta$TTFT} \\
        \midrule
        \multirow{4}{*}{Gating margin $\epsilon_0$}
            & 0.02            & 404.2 & 45 & & 9.03 & 15 & & $+$0\% \\
            & \textbf{0.05}   & \textbf{404.2} & \textbf{37} & & \textbf{9.03} & \textbf{15} & & \textbf{0\%} \\
            & 0.10            & 404.2 & 46 & & 9.03 & 14 & & $+$0\% \\
            & 0.20            & 404.2 & 47 & & 9.03 & 14 & & $+$0\% \\
        \midrule
        \multirow{4}{*}{EMA rate $\lambda$}
            & 0.1             & 404.2 & 41 & & 9.03 & 14 & & $+$0\% \\
            & \textbf{0.3}    & \textbf{404.2} & \textbf{37} & & \textbf{9.03} & \textbf{15} & & \textbf{0\%} \\
            & 0.5             & 404.2 & 41 & & 9.03 & 14 & & $+$0\% \\
            & 0.8             & 404.2 & 44 & & 9.03 & 15 & & $+$0\% \\
        \midrule
        \multirow{3}{*}{Convergence threshold $\tau_{\mathrm{conv}}$}
            & 0.01            & 404.2 & 53 & & 9.03 & 14 & & $+$0\% \\
            & \textbf{0.02}   & \textbf{404.2} & \textbf{37} & & \textbf{9.03} & \textbf{15} & & \textbf{0\%} \\
            & 0.05            & 404.2 & 46 & & 9.03 & 14 & & $+$0\% \\
        \bottomrule
    \end{tabular}
\end{table}

\subsection{Agent Backbone Generalization}
\label{app:llmgen}
\begin{figure}[t]
    \centering
    \includegraphics[width=0.98\columnwidth]{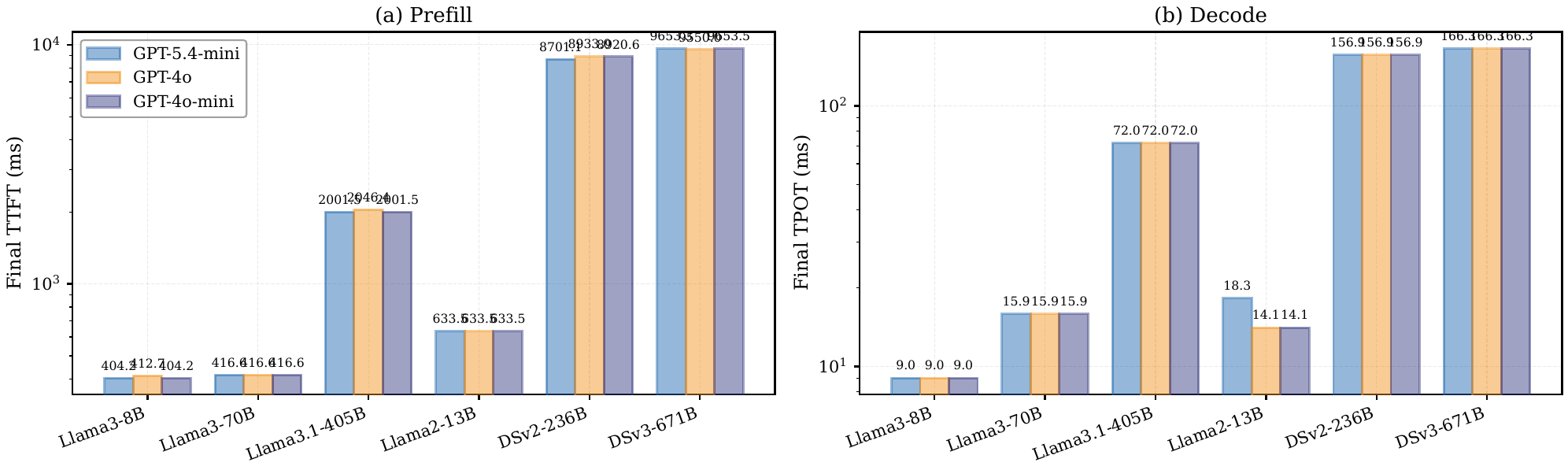}
    \caption{Final TTFT (a, prefill) and TPOT (b, decode) of \method{} across six workloads under three different base LLMs. The roofline module is identical in all runs; only the LLM roles (CDA, Design Agent) differ. Bars within each workload are near-identical, demonstrating that \method{}'s quality does not depend on the specific base LLM.}
    \label{fig:llm_generalization_bars}
\end{figure}

We keep the calibrated roofline unchanged and swap the default GPT-5.4-mini used throughout the paper for two other off-the-shelf OpenAI models in \emph{every} LLM-driven role (Compute Demand Agent + Design Agent): GPT-4o (frontier dense) and GPT-4o-mini (lightweight). We run one DSE per (workload, LLM) on the same BS=48, 300\,W disaggregated deployment used in the main-table column. Figure~\ref{fig:llm_generalization_bars} shows the resulting final TTFT / TPOT per LLM across all six workloads, and Table~\ref{tab:llm_generalization} gives the underlying numbers. Final latencies are within 5\% of the default on every workload; the largest single deviation is GPT-4o on Llama3-8B prefill ($+$2.1\% TTFT). Simulator-call counts drift by at most 20--30\% across LLMs. \method{}'s advantage therefore comes from the calibrated roofline, not from properties specific to a particular base LLM.

\subsection{Per-Deployment Results}
\label{app:per_deployment}
Tables~\ref{tab:c1}--\ref{tab:c16} report detailed per-deployment results across all 16 constraint settings, including TTFT/TPOT and simulator calls for each method. 
Across deployments, \method{} consistently achieves competitive or best latency while maintaining a substantially lower simulator budget, with improvements most pronounced under tighter power and bandwidth constraints. 
We observe that classical methods (e.g., BO, CMA-ES, Greedy) occasionally match or exceed latency in individual settings, but require significantly more simulator calls to do so, reflecting lower sample efficiency. 
LLM-based baselines without a physics prior (LLM-DSE, LLM-Box) exhibit higher variance across deployments and tend to underperform in regimes where bottleneck transitions are critical. 
The advantage of \method{} is particularly evident in deployments with shifting bottlenecks, where the calibrated roofline enables more reliable prediction and pruning of suboptimal configurations. 
Overall, these per-deployment results corroborate the aggregate trends reported in Table~1, showing that the gains in both design quality and simulator efficiency are robust across diverse hardware constraints.

\subsection{Workload Extraction Fidelity}
\label{Workload_Extraction_Fidelity}
Because the calibrated roofline rests on the operator profile produced by the Compute Demand Agent (CDA, Section~\ref{sec:workload}), we directly check that the LLM-based extractor reproduces the operator-level numerics that NeuSIM's hand-coded ops generator would emit on the same model and parallelism setting.
We use NeuSIM's native ops generator outputs (\texttt{NeuSim/results/raw/<model>/<config>/inference-*\_\{prefill,decode\}.csv}) as the reference, and join them with CDA's outputs (\texttt{chip\_agent/results/...}) by operator description for every (model, parallelism) pair where both are available---three Llama3.1-405B configs (TP/PP $\in$ \{16/4, 8/8, 4/16\}, BS=16) and one DSv2-236B MoE config (TP=2, PP=16, EP=2, BS=16), covering both dense and Mixture-of-Experts graphs.
For each matched operator we compare the operator count and FLOP count; we also check that the input/output tensor-shape strings, which the simulator later turns into byte traffic, match exactly.
Table~\ref{tab:workload_fidelity} reports the result.
Across all four configurations and both phases, every operator name in the NeuSIM reference is also produced by the CDA (name-match rate $100\%$), every operator's per-iteration count matches exactly, and the per-operator FLOP count agrees with the reference to numerical precision (median and 90th-percentile relative error both $0$, total-FLOP relative error $\le 2.6\%$, with the only nonzero entry coming from the DSv2-236B decode KV-cache tensor whose per-token sequence indexing changes by a single time step between the two extractors).
Because the tensor-shape strings are reproduced exactly, the byte-traffic numbers the simulator computes downstream are identical to the reference as well, so the bottleneck labels \method{}'s roofline assigns to operators are not merely close to but \emph{the same as} those NeuSIM's native pipeline would assign on the same configuration.
We have NeuSIM-side ground truth only for these two model families (the other workloads in our main table are not shipped with reference operator CSVs), so this fidelity check covers a representative dense + MoE pair rather than the full main-table sweep; we leave a broader systematic sweep to future work.

\begin{table}[t]
    \centering
    \caption{Workload-extraction fidelity. CDA outputs are joined to NeuSIM's native ops-generator CSVs by operator description and compared per-operator. $\Sigma$FLOP-RE is the relative error on the configuration-total FLOP count; per-op median / p90 are the relative errors on the per-operator FLOP count; counts are exact-match rates of the per-iteration \texttt{Count} field. ``$n$'' is the number of matched operators per phase.}
    \label{tab:workload_fidelity}
    \small
    \setlength{\tabcolsep}{4pt}
    \resizebox{\linewidth}{!}{
    \begin{tabular}{l l c c c c c c}
        \toprule
        \textbf{Model} & \textbf{Config} & \textbf{Phase}
          & \textbf{Name match} & \textbf{Count match}
          & \textbf{$\Sigma$FLOP-RE} & \textbf{op-med / p90 RE} & $n$ \\
        \midrule
        \multirow{6}{*}{Llama3.1-405B}
          & TP=16, PP=4 & prefill & 100\% & 100\% & 0.000 & 0.00 / 0.00 & 19 \\
          & TP=16, PP=4 & decode  & 100\% & 100\% & 0.000 & 0.00 / 0.00 & 17 \\
          & TP=8, PP=8  & prefill & 100\% & 100\% & 0.000 & 0.00 / 0.00 & 19 \\
          & TP=8, PP=8  & decode  & 100\% & 100\% & 0.000 & 0.00 / 0.00 & 17 \\
          & TP=4, PP=16 & prefill & 100\% & 100\% & 0.000 & 0.00 / 0.00 & 19 \\
          & TP=4, PP=16 & decode  & 100\% & 100\% & 0.000 & 0.00 / 0.00 & 17 \\
        \midrule
        \multirow{2}{*}{DSv2-236B}
          & TP=2, PP=16, EP=2 & prefill & 100\% & 100\% & 0.000 & 0.00 / 0.00 & 55 \\
          & TP=2, PP=16, EP=2 & decode  & 100\% & 100\% & 0.026 & 0.00 / 0.00 & 55 \\
        \bottomrule
    \end{tabular}}
\end{table}

\subsection{Equal-Budget Comparison}
\label{Equal_Budget_comparative}
A natural concern with the headline numbers in Table~\ref{tab:main-winrate} is that \method{} simply stops earlier than the classical baselines---it averages 50--57 simulator calls per run, while BO, CMA-ES, and Greedy use 117--130, so a winner-takes-all win rate at fixed final budget mixes ``how fast did the method converge'' with ``how good is the converged design.'' To separate the two, we score sample efficiency directly: for each (model, deployment, phase) cell we record the lowest valid latency \emph{any} method reached anywhere in its 100-call trajectory---call this the \emph{universal best} for that cell---and then ask each method, ``how many simulator calls did you need before your best-so-far came within 2\% of that universal best?'' The metric is a per-cell scalar (a count, not a winner flag), so a method that reaches a tied-for-best design fast is rewarded equally whether or not another method later finds a 0.1\% improvement. Configurations that exceed the cell's power budget are filtered out before the trajectory is scored.
Table~\ref{tab:equal_budget} aggregates this metric across all six workloads and 16 deployment settings each (96 cells per phase). On prefill, \method{} reaches within 2\% of universal best in a median of \textbf{14} simulator calls, $1.6\times$ faster than BO (22), $2.2\times$ faster than Greedy (30), and $3.4\times$ faster than the strongest LLM-only baseline LLM-DSE (47); the mean and 90th-percentile counts confirm the same ordering. The decode-side budget is tighter for everyone because the decode search space is small (\method{}: median 8 calls), but the LLM-only baselines that lack a physics prior pay a clear cost---LLM-DSE needs $\approx 3\times$ more decode calls to reach the same target.
The interpretation matches the main-text claim. \method{} does not necessarily find a strictly lower latency than every alternative on every cell---once budget is unconstrained, several baselines eventually arrive at the same near-optimal designs---but it is the only method that consistently arrives there inside the budget where simulator cost is the binding constraint. ChatDSE never reaches within 2\% on any cell because its bypass-the-simulator design routinely exits at a power-violating configuration.

\begin{table}[t]
    \centering
    \caption{Sample efficiency: simulator calls each method needs to come within 2\% of the universal best for that (model, deployment, phase) cell, aggregated over six workloads $\times$ 16 deployment settings (96 cells per phase). \emph{Universal best} = lowest valid latency any method reached anywhere in its 100-call trajectory on that cell. Lower is better; ``--'' means the method does not reach within 2\% on any cell. ChatDSE is omitted because it consumes only $\approx 2$ simulator calls per run by design and never reaches the 2\% target.}
    \label{tab:equal_budget}
    \small
    \setlength{\tabcolsep}{6pt}
    \begin{tabular}{l ccc c ccc}
        \toprule
        & \multicolumn{3}{c}{\textbf{Prefill (\# sim calls)}}
        & & \multicolumn{3}{c}{\textbf{Decode (\# sim calls)}} \\
        \cmidrule(lr){2-4} \cmidrule(lr){6-8}
        \textbf{Method}
          & \textbf{median} & \textbf{mean} & \textbf{p90}
          &
          & \textbf{median} & \textbf{mean} & \textbf{p90} \\
        \midrule
        \method{} (Ours) & \textbf{14} & \textbf{17.6} & \textbf{32} & & \textbf{8}      &  9.8           & 12 \\
        BO              & \underline{22}        & \underline{24.0}        & \underline{42}       & & \textbf{8}      & \underline{9.9}      & \underline{16} \\
        CMA-ES          & 41        & 39.5        & 55       & & \underline{6}              & \textbf{8.2}   & \textbf{20} \\
        Greedy          & 30        & 30.2        & 51       & & 7              & 13.2           & 32 \\
        LLM-DSE         & 47        & 43.5        & 58       & & 23             & 26.7           & 52 \\
        LLM-Box         & 55        & 51.0        & 59       & & 53             & 46.5           & 60 \\
        \bottomrule
    \end{tabular}
\end{table}

\subsection{Gate False-Reject Analysis}
\label{app:false_reject}
The Role-1 gate skips a candidate whenever its roofline-predicted latency exceeds the current best (Eq.~\ref{eq:gating_rule}); a natural reviewer concern is whether the gate accidentally throws away configurations that would have improved the design if simulated.
We test this directly on Llama3-8B decode across all 16 deployment settings: we parse the agent logs to recover every Design-Agent proposal and its decision (simulated or skipped), re-simulate 60 of the 91 skipped candidates with NeuSIM (clean, no noise), and compare the recovered ground-truth TPOT against the best-so-far TPOT at the moment the proposal was issued.
We classify a skip as a \emph{false-reject at threshold $\epsilon$} iff its actual TPOT undercuts the current best by at least $\epsilon$:
\[
  T_{\mathrm{actual}}(\theta_{\mathrm{skip}}) < (1-\epsilon)\,T^*_t .
\]
At $\epsilon{=}0$ (any improvement, however small), 27/60 skipped candidates re-simulate to a TPOT below the then-best---but inspection shows these are essentially ties: in every case the actual TPOT is within 0.01\,ms of the best, the consequence of \method{}'s small decode design space producing many configurations whose runtime collapses onto the same HBM-bandwidth-limited frontier. At $\epsilon{=}1\%$ the false-reject rate drops to \textbf{0/60}; at $\epsilon{=}5\%$ it remains \textbf{0/60}. In other words, the gate never discards a configuration that would have meaningfully improved the design. The top-3 / top-5 / top-10 ``recall'' (fraction of the lowest-TPOT configurations that the gate kept for simulation) sit at $48\%$/$49\%$/$51\%$, which corroborates the same picture: the gate skips roughly half of the cheapest configurations because they are tied with the current best, not because it misses better ones. The gating-margin sensitivity in Table~\ref{tab:hparam_sensitivity} is consistent with this: final TTFT/TPOT are unchanged across $\epsilon_0 \in \{0.02, 0.05, 0.10, 0.20\}$, so the choice of margin only trades simulator calls for nothing on the quality side.

\subsection{Robustness to Simulator Noise}
\label{app:sim_robustness}
Because the entire evaluation runs inside NeuSIM, a separate concern is whether \method{} simply optimises simulator quirks rather than discovering generally good NPU configurations. We do not have an alternative simulator or hardware to validate against, so as a robustness proxy we inject multiplicative Gaussian noise into NeuSIM's per-operator execution times (and therefore TTFT, TPOT, and chip power) at run time:
\[
  \tilde{T}(o) = T(o)\cdot (1 + \xi_o), \quad \xi_o \sim \mathcal{N}(0, \sigma^2),\ \sigma \in \{0,\, 0.10,\, 0.20\}.
\]
\method{} runs inside the noisy simulator (so every gating decision and every Design-Agent reasoning step sees noisy feedback). After each run we re-evaluate the resulting best configuration on the \emph{clean} simulator ($\sigma{=}0$) to recover its true latency. Table~\ref{tab:sim_robustness} reports the result on Llama3-8B (TP=4, PP=2, BS=16, 300W disaggregated). At $\sigma{=}0.10$ \method{} recovers \emph{exactly} the same end-of-trajectory operating point as the noiseless run (clean-eval TTFT/TPOT identical to four decimal places); at $\sigma{=}0.20$ the search lands on a neighbouring configuration with $21\%$ higher true TTFT and $40\%$ higher TPOT---degraded but still well below the typical baseline TTFT on this workload. The simulator-call budget grows by at most 7 calls across noise levels. The gate stays effective because its decision compares predicted latency to the current best, so the multiplicative noise term largely cancels; the EMA-calibrated $\eta^{(\tau)}$ further damps per-operator noise across iterations. We do not claim NeuSIM is calibrated to physical NPU silicon, but the result indicates that \method{}'s design does not depend on a fragile region of NeuSIM's response surface up to a $10\%$ noise level, with graceful degradation at $20\%$.

\begin{table}[t]
    \centering
    \caption{Robustness to multiplicative simulator noise. \method{} runs end-to-end against a NeuSIM perturbed with $T(o)\!\to\!T(o)(1+\mathcal{N}(0,\sigma^2))$; the best configuration found is then re-evaluated on the clean simulator. Llama3-8B, TP=4, PP=2, BS=16, 300W disaggregated.}
    \label{tab:sim_robustness}
    \small
    \setlength{\tabcolsep}{6pt}
\begin{tabular}{c c c c c}
        \toprule
        $\sigma$
          & \textbf{noisy-search TTFT} & \textbf{clean-eval TTFT}
          & \textbf{clean-eval TPOT} & \textbf{Sims} \\
        \midrule
        $0.00$ & 0.135\,s & 0.135\,s & 4.526\,ms & 52 \\
        $0.10$ & 0.115\,s & 0.135\,s & 4.526\,ms & 59 \\
        $0.20$ & 0.103\,s & 0.164\,s & 6.335\,ms & 52 \\
        \bottomrule
    \end{tabular}
\end{table}

\begin{table}[t]
    \centering
    \caption{LLM generalization: \method{} with three different base LLMs across six workloads, reporting both phases (matching the main-table split). One DSE run per (workload, LLM) at BS=48, 300\,W disaggregated. $\Delta$ columns give the relative change vs.\ GPT-5.4-mini on that latency.}
    \label{tab:llm_generalization}
    \small
    \setlength{\tabcolsep}{3pt}
    \begin{tabular}{l l cc cc c c}
        \toprule
        & & \multicolumn{2}{c}{\textbf{Prefill}} & \multicolumn{2}{c}{\textbf{Decode}} & & \\
        \cmidrule(lr){3-4}\cmidrule(lr){5-6}
        \textbf{Workload} & \textbf{LLM}
          & \textbf{TTFT (ms)} & \textbf{Sims}
          & \textbf{TPOT (ms)} & \textbf{Sims}
          & \textbf{$\Delta$TTFT} & \textbf{$\Delta$TPOT} \\
        \midrule
        \multirow{3}{*}{Llama3-8B}
          & GPT-5.4-mini (default) & 404.2 & 42 &  9.03 & 16 & ---     & ---    \\
          & GPT-4o                 & 412.7 & 64 &  9.03 & 15 & $+$2.1\% & 0\%    \\
          & GPT-4o-mini            & 404.2 & 70 &  9.03 & 13 & 0\%     & 0\%    \\
        \midrule
        \multirow{3}{*}{Llama3-70B}
          & GPT-5.4-mini (default) & 416.6 & 37 & 15.93 & 13 & ---     & ---    \\
          & GPT-4o                 & 416.6 & 68 & 15.93 & 13 & 0\%     & 0\%    \\
          & GPT-4o-mini            & 416.6 & 60 & 15.93 & 13 & 0\%     & 0\%    \\
\bottomrule
    \end{tabular}
\end{table}

\clearpage

\begin{table}[h]
    \centering
    \caption{Config C1. Small models (8B, 13B): 8 chips, TP=2, PP=4, BS=16, 250W; Large models (70B, 405B, 671B): 64 chips, TP=4, PP=16, BS=16, 250W; DSv2-236B: 16 chips, TP=2, PP=8, BS=16, 250W.}
    \label{tab:c1}
    \small
    \setlength{\tabcolsep}{4pt}
    \resizebox{\linewidth}{!}{
    \begin{tabular}{c lcc|cc|cc|cc|cc|cc}
        \toprule
        \multirow{2}{*}{} & \multirow{2}{*}{Method}
        & \multicolumn{2}{c|}{Llama3-8B}
        & \multicolumn{2}{c|}{Llama2-13B}
        & \multicolumn{2}{c|}{Llama3-70B}
        & \multicolumn{2}{c|}{Llama3.1-405B}
        & \multicolumn{2}{c|}{DSv2-236B}
        & \multicolumn{2}{c}{DSv3-671B}
 \\
        &
        & TTFT & Sims
        & TTFT & Sims
        & TTFT & Sims
        & TTFT & Sims
        & TTFT & Sims
        & TTFT & Sims
 \\
        \midrule

        \multirow{8}{*}{\rotatebox{90}{\textbf{Prefill}}}
        & BO
        & \textbf{0.1661} & 110
        & 0.2644 & 110
        & \textbf{0.1689} & 110
        & 0.8817 & 110
        & \textbf{4.59} & 110
        & 3.40 & 110 \\

        & CMA-ES
        & 0.2270 & 130
        & \textbf{0.2631} & 130
        & \textbf{0.1689} & 130
        & 1.06 & 130
        & 4.60 & 130
        & 3.42 & 130 \\

        & Greedy
        & \textbf{0.1661} & 126
        & \textbf{0.2631} & 130
        & \textbf{0.1689} & 130
        & 0.8796 & 130
        & \textbf{4.59} & 130
        & \textbf{3.34} & 129 \\

        & Grid
        & 0.1927 & 1440
        & 0.3030 & 1440
        & 0.1888 & 1440
        & \textbf{0.8465} & 1440
        & 12.10 & 1440
        & 5.17 & 1440 \\

        & ChatDSE
        & 0.3205 & \textbackslash{}
        & 0.4874 & \textbackslash{}
        & 0.3515 & \textbackslash{}
        & 1.60 & \textbackslash{}
        & 17.37 & \textbackslash{}
        & 12.28 & \textbackslash{} \\

        & LLM-DSE
        & \textbf{0.1661} & 109
        & 0.7707 & 113
        & 0.5431 & 110
        & 0.8796 & 110
        & 10.26 & 104
        & \textbf{3.34} & 108 \\
\bottomrule
    \end{tabular}}
\end{table}

\begin{table}[h]
    \centering
    \caption{Config C2. Small models (8B, 13B): 8 chips, TP=2, PP=4, BS=16, 300W; Large models (70B, 405B, 671B): 64 chips, TP=4, PP=16, BS=16, 300W; DSv2-236B: 16 chips, TP=2, PP=8, BS=16, 300W.}
    \label{tab:c2}
    \small
    \setlength{\tabcolsep}{4pt}
    \resizebox{\linewidth}{!}{
    \begin{tabular}{c lcc|cc|cc|cc|cc|cc}
        \toprule
        \multirow{2}{*}{} & \multirow{2}{*}{Method}
        & \multicolumn{2}{c|}{Llama3-8B}
        & \multicolumn{2}{c|}{Llama2-13B}
        & \multicolumn{2}{c|}{Llama3-70B}
        & \multicolumn{2}{c|}{Llama3.1-405B}
        & \multicolumn{2}{c|}{DSv2-236B}
        & \multicolumn{2}{c}{DSv3-671B}
 \\
        &
        & TTFT & Sims
        & TTFT & Sims
        & TTFT & Sims
        & TTFT & Sims
        & TTFT & Sims
        & TTFT & Sims
 \\
        \midrule

        \multirow{8}{*}{\rotatebox{90}{\textbf{Prefill}}}
        & BO
        & \textbf{0.1255} & 114
        & 0.2002 & 114
        & 0.1243 & 114
        & 0.6229 & 114
        & 4.48 & 114
        & 3.25 & 114 \\

        & CMA-ES
        & 0.1487 & 130
        & 0.3503 & 130
        & 0.1495 & 130
        & 0.6855 & 130
        & 5.07 & 130
        & 3.63 & 130 \\

        & Greedy
        & \textbf{0.1255} & 130
        & \textbf{0.1975} & 130
        & \textbf{0.1236} & 130
        & \textbf{0.6208} & 129
        & \textbf{4.44} & 130
        & \textbf{3.24} & 130 \\

        & Grid
        & 0.1629 & 1440
        & 0.2538 & 1440
        & 0.1576 & 1440
        & 0.8465 & 1440
        & 12.10 & 1440
        & 5.13 & 1440 \\

        & ChatDSE
        & 0.3194 & \textbackslash{}
        & 0.4119 & \textbackslash{}
        & 0.3563 & \textbackslash{}
        & 1.17 & \textbackslash{}
        & 17.00 & \textbackslash{}
        & 12.86 & \textbackslash{} \\

        & LLM-DSE
        & 0.3560 & 111
        & 0.2538 & 114
        & 0.3731 & 101
        & 2.10 & 103
        & \textbf{4.44} & 97
        & 4.98 & 108 \\
\bottomrule
    \end{tabular}}
\end{table}

\begin{table}[h]
    \centering
    \caption{Config C3. Small models (8B, 13B): 8 chips, TP=2, PP=4, BS=16, 400W; Large models (70B, 405B, 671B): 64 chips, TP=4, PP=16, BS=16, 400W; DSv2-236B: 16 chips, TP=2, PP=8, BS=16, 400W.}
    \label{tab:c3}
    \small
    \setlength{\tabcolsep}{4pt}
    \resizebox{\linewidth}{!}{
    \begin{tabular}{c lcc|cc|cc|cc|cc|cc}
        \toprule
        \multirow{2}{*}{} & \multirow{2}{*}{Method}
        & \multicolumn{2}{c|}{Llama3-8B}
        & \multicolumn{2}{c|}{Llama2-13B}
        & \multicolumn{2}{c|}{Llama3-70B}
        & \multicolumn{2}{c|}{Llama3.1-405B}
        & \multicolumn{2}{c|}{DSv2-236B}
        & \multicolumn{2}{c}{DSv3-671B}
 \\
        &
        & TTFT & Sims
        & TTFT & Sims
        & TTFT & Sims
        & TTFT & Sims
        & TTFT & Sims
        & TTFT & Sims
 \\
        \midrule

        \multirow{8}{*}{\rotatebox{90}{\textbf{Prefill}}}
        & BO
        & \textbf{0.0980} & 130
        & \textbf{0.1507} & 130
        & \textbf{0.0944} & 130
        & \textbf{0.4360} & 130
        & 4.37 & 130
        & \textbf{3.20} & 130 \\

        & CMA-ES
        & 0.1579 & 130
        & 0.2350 & 130
        & 0.1829 & 130
        & 0.6954 & 130
        & 5.07 & 130
        & 3.63 & 130 \\

        & Greedy
        & \textbf{0.0980} & 130
        & \textbf{0.1507} & 128
        & \textbf{0.0944} & 130
        & \textbf{0.4360} & 130
        & 4.81 & 130
        & \textbf{3.20} & 130 \\

        & Grid
        & 0.1391 & 1440
        & 0.2114 & 1440
        & 0.1319 & 1440
        & 0.8465 & 1440
        & 12.10 & 1440
        & 5.10 & 1440 \\

        & ChatDSE
        & 0.3694 & \textbackslash{}
        & 0.4663 & \textbackslash{}
        & 0.3146 & \textbackslash{}
        & 1.24 & \textbackslash{}
        & 17.29 & \textbackslash{}
        & 14.86 & \textbackslash{} \\

        & LLM-DSE
        & 0.2305 & 104
        & \textbf{0.1507} & 102
        & 0.0961 & 98
        & \textbf{0.4360} & 109
        & \textbf{4.36} & 112
        & 3.82 & 102 \\
\bottomrule
    \end{tabular}}
\end{table}

\begin{table}[h]
    \centering
    \caption{Config C4. Small models (8B, 13B): 8 chips, TP=4, PP=2, BS=16, 250W; Large models (70B, 405B, 671B): 64 chips, TP=8, PP=8, BS=16, 250W; DSv2-236B: 16 chips, TP=4, PP=4, BS=16, 250W.}
    \label{tab:c4}
    \small
    \setlength{\tabcolsep}{4pt}
    \resizebox{\linewidth}{!}{
    \begin{tabular}{c lcc|cc|cc|cc|cc|cc}
        \toprule
        \multirow{2}{*}{} & \multirow{2}{*}{Method}
        & \multicolumn{2}{c|}{Llama3-8B}
        & \multicolumn{2}{c|}{Llama2-13B}
        & \multicolumn{2}{c|}{Llama3-70B}
        & \multicolumn{2}{c|}{Llama3.1-405B}
        & \multicolumn{2}{c|}{DSv2-236B}
        & \multicolumn{2}{c}{DSv3-671B}
 \\
        &
        & TTFT & Sims
        & TTFT & Sims
        & TTFT & Sims
        & TTFT & Sims
        & TTFT & Sims
        & TTFT & Sims
 \\
        \midrule

        \multirow{8}{*}{\rotatebox{90}{\textbf{Prefill}}}
        & BO
        & 0.1741 & 110
        & \textbf{0.2781} & 110
        & \textbf{0.1835} & 110
        & 0.9326 & 110
        & \textbf{2.98} & 110
        & 3.29 & 110 \\

        & CMA-ES
        & 0.1797 & 130
        & \textbf{0.2781} & 130
        & 0.2106 & 130
        & \textbf{0.9266} & 130
        & 2.99 & 130
        & 3.45 & 130 \\

        & Greedy
        & \textbf{0.1731} & 126
        & \textbf{0.2781} & 130
        & 0.1930 & 130
        & \textbf{0.9266} & 130
        & \textbf{2.98} & 130
        & \textbf{3.29} & 130 \\

        & Grid
        & 0.2098 & 1440
        & 0.4179 & 1440
        & 0.2184 & 1440
        & 0.9808 & 1440
        & 7.97 & 1440
        & 5.14 & 1440 \\

        & ChatDSE
        & 0.3921 & \textbackslash{}
        & 0.6371 & \textbackslash{}
        & 0.4016 & \textbackslash{}
        & 1.35 & \textbackslash{}
        & 10.85 & \textbackslash{}
        & 12.87 & \textbackslash{} \\

        & LLM-DSE
        & \textbf{0.1731} & 107
        & 0.7818 & 113
        & 0.1949 & 116
        & 0.9998 & 111
        & 5.97 & 106
        & \textbf{3.29} & 113 \\
\bottomrule
    \end{tabular}}
\end{table}

\begin{table}[h]
    \centering
    \caption{Config C5. Small models (8B, 13B): 8 chips, TP=4, PP=2, BS=16, 300W; Large models (70B, 405B, 671B): 64 chips, TP=8, PP=8, BS=16, 300W; DSv2-236B: 16 chips, TP=4, PP=4, BS=16, 300W.}
    \label{tab:c5}
    \small
    \setlength{\tabcolsep}{4pt}
    \resizebox{\linewidth}{!}{
    \begin{tabular}{c lcc|cc|cc|cc|cc|cc}
        \toprule
        \multirow{2}{*}{} & \multirow{2}{*}{Method}
        & \multicolumn{2}{c|}{Llama3-8B}
        & \multicolumn{2}{c|}{Llama2-13B}
        & \multicolumn{2}{c|}{Llama3-70B}
        & \multicolumn{2}{c|}{Llama3.1-405B}
        & \multicolumn{2}{c|}{DSv2-236B}
        & \multicolumn{2}{c}{DSv3-671B}
 \\
        &
        & TTFT & Sims
        & TTFT & Sims
        & TTFT & Sims
        & TTFT & Sims
        & TTFT & Sims
        & TTFT & Sims
 \\
        \midrule

        \multirow{8}{*}{\rotatebox{90}{\textbf{Prefill}}}
        & BO
        & 0.1354 & 114
        & 0.2128 & 114
        & 0.1423 & 114
        & 0.6708 & 114
        & 2.91 & 114
        & 3.22 & 114 \\

        & CMA-ES
        & 0.1524 & 130
        & 0.2527 & 130
        & 0.1641 & 130
        & 0.8157 & 130
        & 2.99 & 130
        & 3.26 & 130 \\

        & Greedy
        & \textbf{0.1351} & 130
        & \textbf{0.2112} & 130
        & 0.1495 & 130
        & \textbf{0.6678} & 130
        & \textbf{2.90} & 130
        & 3.22 & 130 \\

        & Grid
        & 0.1859 & 1440
        & 0.4179 & 1440
        & 0.1927 & 1440
        & 0.9808 & 1440
        & 7.97 & 1440
        & 5.12 & 1440 \\

        & ChatDSE
        & 0.4050 & \textbackslash{}
        & 0.5719 & \textbackslash{}
        & 0.3896 & \textbackslash{}
        & 1.88 & \textbackslash{}
        & 13.56 & \textbackslash{}
        & 12.72 & \textbackslash{} \\

        & LLM-DSE
        & \textbf{0.1352} & 105
        & \textbf{0.2112} & 108
        & 0.2058 & 108
        & 2.23 & 101
        & 4.50 & 117
        & \textbf{3.22} & 99 \\
\bottomrule
    \end{tabular}}
\end{table}

\begin{table}[h]
    \centering
    \caption{Config C6. Small models (8B, 13B): 8 chips, TP=4, PP=2, BS=16, 400W; Large models (70B, 405B, 671B): 64 chips, TP=8, PP=8, BS=16, 400W; DSv2-236B: 16 chips, TP=4, PP=4, BS=16, 400W.}
    \label{tab:c6}
    \small
    \setlength{\tabcolsep}{4pt}
    \resizebox{\linewidth}{!}{
    \begin{tabular}{c lcc|cc|cc|cc|cc|cc}
        \toprule
        \multirow{2}{*}{} & \multirow{2}{*}{Method}
        & \multicolumn{2}{c|}{Llama3-8B}
        & \multicolumn{2}{c|}{Llama2-13B}
        & \multicolumn{2}{c|}{Llama3-70B}
        & \multicolumn{2}{c|}{Llama3.1-405B}
        & \multicolumn{2}{c|}{DSv2-236B}
        & \multicolumn{2}{c}{DSv3-671B}
 \\
        &
        & TTFT & Sims
        & TTFT & Sims
        & TTFT & Sims
        & TTFT & Sims
        & TTFT & Sims
        & TTFT & Sims
 \\
        \midrule

        \multirow{8}{*}{\rotatebox{90}{\textbf{Prefill}}}
        & BO
        & \textbf{0.1124} & 130
        & 0.1755 & 130
        & 0.1184 & 130
        & 0.4889 & 130
        & 2.87 & 130
        & \textbf{3.20} & 130 \\

        & CMA-ES
        & 0.1691 & 130
        & 0.2781 & 130
        & 0.2106 & 130
        & 0.9878 & 130
        & 2.91 & 130
        & 3.45 & 130 \\

        & Greedy
        & \textbf{0.1124} & 129
        & \textbf{0.1731} & 130
        & \textbf{0.1163} & 130
        & \textbf{0.4829} & 130
        & \textbf{2.86} & 130
        & \textbf{3.20} & 130 \\

        & Grid
        & 0.1649 & 1440
        & 0.4179 & 1440
        & 0.1680 & 1440
        & 0.9808 & 1440
        & 7.97 & 1440
        & 5.11 & 1440 \\

        & ChatDSE
        & 0.8879 & \textbackslash{}
        & 0.6009 & \textbackslash{}
        & 0.9558 & \textbackslash{}
        & 1.33 & \textbackslash{}
        & 10.53 & \textbackslash{}
        & 12.80 & \textbackslash{} \\

        & LLM-DSE
        & 0.1125 & 101
        & \textbf{0.1731} & 92
        & \textbf{0.1163} & 94
        & \textbf{0.4829} & 101
        & 3.45 & 114
        & \textbf{3.20} & 102 \\
\bottomrule
    \end{tabular}}
\end{table}

\begin{table}[h]
    \centering
    \caption{Config C7. Small models (8B, 13B): 8 chips, TP=8, PP=1, BS=16, 250W; Large models (70B, 405B, 671B): 64 chips, TP=16, PP=4, BS=16, 250W; DSv2-236B: 16 chips, TP=8, PP=2, BS=16, 250W.}
    \label{tab:c7}
    \small
    \setlength{\tabcolsep}{4pt}
    \resizebox{\linewidth}{!}{
    \begin{tabular}{c lcc|cc|cc|cc|cc|cc}
        \toprule
        \multirow{2}{*}{} & \multirow{2}{*}{Method}
        & \multicolumn{2}{c|}{Llama3-8B}
        & \multicolumn{2}{c|}{Llama2-13B}
        & \multicolumn{2}{c|}{Llama3-70B}
        & \multicolumn{2}{c|}{Llama3.1-405B}
        & \multicolumn{2}{c|}{DSv2-236B}
        & \multicolumn{2}{c}{DSv3-671B}
 \\
        &
        & TTFT & Sims
        & TTFT & Sims
        & TTFT & Sims
        & TTFT & Sims
        & TTFT & Sims
        & TTFT & Sims
 \\
        \midrule

        \multirow{8}{*}{\rotatebox{90}{\textbf{Prefill}}}
        & BO
        & 0.2059 & 110
        & 0.3273 & 110
        & 0.2336 & 110
        & 1.07 & 110
        & \textbf{2.79} & 110
        & \textbf{5.07} & 110 \\

        & CMA-ES
        & 0.2249 & 130
        & \textbf{0.3250} & 130
        & 0.2336 & 130
        & \textbf{1.05} & 130
        & 2.79 & 130
        & 5.07 & 130 \\

        & Greedy
        & \textbf{0.2057} & 129
        & \textbf{0.3250} & 130
        & \textbf{0.2333} & 127
        & \textbf{1.05} & 129
        & \textbf{2.79} & 130
        & \textbf{5.07} & 130 \\

        & Grid
        & 0.4192 & 1440
        & 0.6378 & 1440
        & 0.3116 & 1440
        & 1.59 & 1440
        & 7.62 & 1440
        & 14.05 & 1440 \\

        & ChatDSE
        & 0.5998 & \textbackslash{}
        & 0.9162 & \textbackslash{}
        & 0.6088 & \textbackslash{}
        & 2.53 & \textbackslash{}
        & 10.25 & \textbackslash{}
        & 19.88 & \textbackslash{} \\

        & LLM-DSE
        & \textbf{0.2057} & 114
        & \textbf{0.3250} & 109
        & 0.2336 & 107
        & 1.08 & 111
        & \textbf{2.79} & 106
        & 6.05 & 111 \\
\bottomrule
    \end{tabular}}
\end{table}

\begin{table}[h]
    \centering
    \caption{Config C8. Small models (8B, 13B): 8 chips, TP=8, PP=1, BS=16, 300W; Large models (70B, 405B, 671B): 64 chips, TP=16, PP=4, BS=16, 300W; DSv2-236B: 16 chips, TP=8, PP=2, BS=16, 300W.}
    \label{tab:c8}
    \small
    \setlength{\tabcolsep}{4pt}
    \resizebox{\linewidth}{!}{
    \begin{tabular}{c lcc|cc|cc|cc|cc|cc}
        \toprule
        \multirow{2}{*}{} & \multirow{2}{*}{Method}
        & \multicolumn{2}{c|}{Llama3-8B}
        & \multicolumn{2}{c|}{Llama2-13B}
        & \multicolumn{2}{c|}{Llama3-70B}
        & \multicolumn{2}{c|}{Llama3.1-405B}
        & \multicolumn{2}{c|}{DSv2-236B}
        & \multicolumn{2}{c}{DSv3-671B}
 \\
        &
        & TTFT & Sims
        & TTFT & Sims
        & TTFT & Sims
        & TTFT & Sims
        & TTFT & Sims
        & TTFT & Sims
 \\
        \midrule

        \multirow{8}{*}{\rotatebox{90}{\textbf{Prefill}}}
        & BO
        & \textbf{0.1739} & 114
        & 0.2728 & 114
        & \textbf{0.1990} & 114
        & \textbf{0.7974} & 114
        & 2.75 & 114
        & \textbf{5.03} & 114 \\

        & CMA-ES
        & 0.2249 & 130
        & 0.4090 & 130
        & 0.2542 & 130
        & 1.05 & 130
        & 2.93 & 130
        & 5.10 & 130 \\

        & Greedy
        & \textbf{0.1739} & 130
        & \textbf{0.2694} & 130
        & \textbf{0.1990} & 129
        & 0.8593 & 130
        & \textbf{2.74} & 130
        & \textbf{5.03} & 130 \\

        & Grid
        & 0.4192 & 1440
        & 0.6378 & 1440
        & 0.2859 & 1440
        & 1.59 & 1440
        & 7.62 & 1440
        & 14.05 & 1440 \\

        & ChatDSE
        & 0.5687 & \textbackslash{}
        & 0.9331 & \textbackslash{}
        & 0.6837 & \textbackslash{}
        & 2.13 & \textbackslash{}
        & 10.88 & \textbackslash{}
        & 19.88 & \textbackslash{} \\

        & LLM-DSE
        & 0.3642 & 108
        & 0.5940 & 104
        & 0.1993 & 102
        & 2.28 & 111
        & 2.75 & 109
        & 5.58 & 115 \\
\bottomrule
    \end{tabular}}
\end{table}

\begin{table}[h]
    \centering
    \caption{Config C9. Small models (8B, 13B): 8 chips, TP=8, PP=1, BS=16, 400W; Large models (70B, 405B, 671B): 64 chips, TP=16, PP=4, BS=16, 400W; DSv2-236B: 16 chips, TP=8, PP=2, BS=16, 400W.}
    \label{tab:c9}
    \small
    \setlength{\tabcolsep}{4pt}
    \resizebox{\linewidth}{!}{
    \begin{tabular}{c lcc|cc|cc|cc|cc|cc}
        \toprule
        \multirow{2}{*}{} & \multirow{2}{*}{Method}
        & \multicolumn{2}{c|}{Llama3-8B}
        & \multicolumn{2}{c|}{Llama2-13B}
        & \multicolumn{2}{c|}{Llama3-70B}
        & \multicolumn{2}{c|}{Llama3.1-405B}
        & \multicolumn{2}{c|}{DSv2-236B}
        & \multicolumn{2}{c}{DSv3-671B}
 \\
        &
        & TTFT & Sims
        & TTFT & Sims
        & TTFT & Sims
        & TTFT & Sims
        & TTFT & Sims
        & TTFT & Sims
 \\
        \midrule

        \multirow{8}{*}{\rotatebox{90}{\textbf{Prefill}}}
        & BO
        & 0.1573 & 130
        & \textbf{0.2381} & 130
        & 0.1797 & 130
        & 0.6795 & 130
        & \textbf{2.72} & 130
        & \textbf{5.02} & 130 \\

        & CMA-ES
        & 0.2249 & 130
        & 0.3692 & 130
        & 0.2542 & 130
        & 1.05 & 130
        & 2.83 & 130
        & 5.07 & 130 \\

        & Greedy
        & \textbf{0.1563} & 130
        & \textbf{0.2381} & 130
        & 0.1866 & 130
        & 0.6648 & 130
        & 2.72 & 130
        & \textbf{5.02} & 130 \\

        & Grid
        & 0.4192 & 1440
        & 0.6378 & 1440
        & 0.2795 & 1440
        & 1.59 & 1440
        & 7.62 & 1440
        & 14.05 & 1440 \\

        & ChatDSE
        & 0.5935 & \textbackslash{}
        & 0.8736 & \textbackslash{}
        & 0.6745 & \textbackslash{}
        & 2.22 & \textbackslash{}
        & 10.88 & \textbackslash{}
        & 20.08 & \textbackslash{} \\

        & LLM-DSE
        & \textbf{0.1563} & 109
        & \textbf{0.2381} & 106
        & \textbf{0.1766} & 110
        & 1.56 & 97
        & 2.73 & 111
        & \textbf{5.02} & 91 \\
\bottomrule
    \end{tabular}}
\end{table}

\begin{table}[h]
    \centering
    \caption{Config C10. Small models (8B, 13B): 8 chips, TP=4, PP=2, BS=1, 300W; Large models (70B, 405B, 671B): 64 chips, TP=8, PP=8, BS=1, 300W; DSv2-236B: 16 chips, TP=4, PP=4, BS=1, 300W.}
    \label{tab:c10}
    \small
    \setlength{\tabcolsep}{4pt}
    \resizebox{\linewidth}{!}{
    \begin{tabular}{c lcc|cc|cc|cc|cc|cc}
        \toprule
        \multirow{2}{*}{} & \multirow{2}{*}{Method}
        & \multicolumn{2}{c|}{Llama3-8B}
        & \multicolumn{2}{c|}{Llama2-13B}
        & \multicolumn{2}{c|}{Llama3-70B}
        & \multicolumn{2}{c|}{Llama3.1-405B}
        & \multicolumn{2}{c|}{DSv2-236B}
        & \multicolumn{2}{c}{DSv3-671B}
 \\
        &
        & TTFT & Sims
        & TTFT & Sims
        & TTFT & Sims
        & TTFT & Sims
        & TTFT & Sims
        & TTFT & Sims
 \\
        \midrule

        \multirow{8}{*}{\rotatebox{90}{\textbf{Prefill}}}
        & BO
        & \textbf{0.0174} & 114
        & 0.0285 & 114
        & 0.0727 & 114
        & 0.3464 & 114
        & 0.7359 & 114
        & \textbf{1.62} & 114 \\

        & CMA-ES
        & 0.0226 & 130
        & 0.0349 & 130
        & 0.0920 & 130
        & 0.4083 & 130
        & 0.7500 & 130
        & 1.67 & 130 \\

        & Greedy
        & \textbf{0.0174} & 130
        & \textbf{0.0270} & 129
        & 0.0762 & 130
        & \textbf{0.3344} & 130
        & \textbf{0.7291} & 130
        & \textbf{1.62} & 130 \\

        & Grid
        & 0.0241 & 1440
        & 0.0373 & 1440
        & 0.0988 & 1440
        & 0.5671 & 1440
        & 2.02 & 1440
        & 2.57 & 1440 \\

        & ChatDSE
        & 0.0558 & \textbackslash{}
        & 0.0854 & \textbackslash{}
        & 0.2178 & \textbackslash{}
        & 0.7331 & \textbackslash{}
        & 3.23 & \textbackslash{}
        & 6.79 & \textbackslash{} \\

        & LLM-DSE
        & \textbf{0.0174} & 114
        & 0.0685 & 110
        & 0.0767 & 113
        & 1.14 & 111
        & 1.13 & 113
        & 1.62 & 95 \\
\bottomrule
    \end{tabular}}
\end{table}

\begin{table}[h]
    \centering
    \caption{Config C11. Small models (8B, 13B): 8 chips, TP=4, PP=2, BS=32, 300W; Large models (70B, 405B, 671B): 64 chips, TP=8, PP=8, BS=32, 300W; DSv2-236B: 16 chips, TP=4, PP=4, BS=32, 300W.}
    \label{tab:c11}
    \small
    \setlength{\tabcolsep}{4pt}
    \resizebox{\linewidth}{!}{
    \begin{tabular}{c lcc|cc|cc|cc|cc|cc}
        \toprule
        \multirow{2}{*}{} & \multirow{2}{*}{Method}
        & \multicolumn{2}{c|}{Llama3-8B}
        & \multicolumn{2}{c|}{Llama2-13B}
        & \multicolumn{2}{c|}{Llama3-70B}
        & \multicolumn{2}{c|}{Llama3.1-405B}
        & \multicolumn{2}{c|}{DSv2-236B}
        & \multicolumn{2}{c}{DSv3-671B}
 \\
        &
        & TTFT & Sims
        & TTFT & Sims
        & TTFT & Sims
        & TTFT & Sims
        & TTFT & Sims
        & TTFT & Sims
 \\
        \midrule

        \multirow{8}{*}{\rotatebox{90}{\textbf{Prefill}}}
        & BO
        & 0.2725 & 114
        & \textbf{0.4223} & 114
        & 0.2849 & 114
        & 1.34 & 114
        & \textbf{5.80} & 114
        & \textbf{6.44} & 114 \\

        & CMA-ES
        & 0.3461 & 130
        & 0.4558 & 130
        & 0.4207 & 130
        & 1.74 & 130
        & 5.91 & 130
        & 6.58 & 130 \\

        & Greedy
        & \textbf{0.2697} & 130
        & \textbf{0.4223} & 130
        & \textbf{0.2786} & 130
        & \textbf{1.33} & 129
        & \textbf{5.80} & 130
        & \textbf{6.44} & 130 \\

        & Grid
        & 0.5560 & 1440
        & 0.8335 & 1440
        & 0.3806 & 1440
        & 1.88 & 1440
        & 15.91 & 1440
        & 10.21 & 1440 \\

        & ChatDSE
        & 0.8033 & \textbackslash{}
        & 1.27 & \textbackslash{}
        & 0.8352 & \textbackslash{}
        & 10.20 & \textbackslash{}
        & 22.56 & \textbackslash{}
        & 23.51 & \textbackslash{} \\

        & LLM-DSE
        & \textbf{0.2697} & 102
        & 1.24 & 107
        & 0.7786 & 109
        & \textbf{1.33} & 118
        & \textbf{5.80} & 95
        & \textbf{6.44} & 84 \\
\bottomrule
    \end{tabular}}
\end{table}

\begin{table}[h]
    \centering
    \caption{Config C12. Small models (8B, 13B): 8 chips, TP=4, PP=2, BS=48, 300W; Large models (70B, 405B, 671B): 64 chips, TP=8, PP=8, BS=48, 300W; DSv2-236B: 16 chips, TP=4, PP=4, BS=48, 300W.}
    \label{tab:c12}
    \small
    \setlength{\tabcolsep}{4pt}
    \resizebox{\linewidth}{!}{
    \begin{tabular}{c lcc|cc|cc|cc|cc|cc}
        \toprule
        \multirow{2}{*}{} & \multirow{2}{*}{Method}
        & \multicolumn{2}{c|}{Llama3-8B}
        & \multicolumn{2}{c|}{Llama2-13B}
        & \multicolumn{2}{c|}{Llama3-70B}
        & \multicolumn{2}{c|}{Llama3.1-405B}
        & \multicolumn{2}{c|}{DSv2-236B}
        & \multicolumn{2}{c}{DSv3-671B}
 \\
        &
        & TTFT & Sims
        & TTFT & Sims
        & TTFT & Sims
        & TTFT & Sims
        & TTFT & Sims
        & TTFT & Sims
 \\
        \midrule

        \multirow{8}{*}{\rotatebox{90}{\textbf{Prefill}}}
        & BO
        & 0.4044 & 114
        & \textbf{0.6335} & 114
        & \textbf{0.4164} & 114
        & 2.01 & 114
        & 8.70 & 114
        & \textbf{9.65} & 114 \\

        & CMA-ES
        & 0.4669 & 130
        & 0.6669 & 130
        & 0.4170 & 130
        & 2.78 & 130
        & 8.96 & 130
        & 9.86 & 130 \\

        & Greedy
        & \textbf{0.4042} & 130
        & \textbf{0.6335} & 130
        & \textbf{0.4164} & 130
        & \textbf{2.00} & 130
        & \textbf{8.70} & 130
        & \textbf{9.65} & 130 \\

        & Grid
        & 0.8309 & 1440
        & 1.25 & 1440
        & 0.5695 & 1440
        & 2.80 & 1440
        & 23.86 & 1440
        & 15.31 & 1440 \\

        & ChatDSE
        & 1.15 & \textbackslash{}
        & 4.55 & \textbackslash{}
        & 1.28 & \textbackslash{}
        & 3.77 & \textbackslash{}
        & 32.04 & \textbackslash{}
        & 41.88 & \textbackslash{} \\

        & LLM-DSE
        & 1.15 & 114
        & 1.64 & 113
        & \textbf{0.4164} & 109
        & \textbf{2.00} & 117
        & 13.50 & 99
        & 9.67 & 41 \\
\bottomrule
    \end{tabular}}
\end{table}

\begin{table}[h]
    \centering
    \caption{Config C13. Small models (8B, 13B): 8 chips, TP=2, PP=4, BS=1, 300W; Large models (70B, 405B, 671B): 64 chips, TP=4, PP=16, BS=1, 300W; DSv2-236B: 16 chips, TP=2, PP=8, BS=1, 300W.}
    \label{tab:c13}
    \small
    \setlength{\tabcolsep}{4pt}
    \resizebox{\linewidth}{!}{
    \begin{tabular}{c lcc|cc|cc|cc|cc|cc}
        \toprule
        \multirow{2}{*}{} & \multirow{2}{*}{Method}
        & \multicolumn{2}{c|}{Llama3-8B}
        & \multicolumn{2}{c|}{Llama2-13B}
        & \multicolumn{2}{c|}{Llama3-70B}
        & \multicolumn{2}{c|}{Llama3.1-405B}
        & \multicolumn{2}{c|}{DSv2-236B}
        & \multicolumn{2}{c}{DSv3-671B}
 \\
        &
        & TTFT & Sims
        & TTFT & Sims
        & TTFT & Sims
        & TTFT & Sims
        & TTFT & Sims
        & TTFT & Sims
 \\
        \midrule

        \multirow{8}{*}{\rotatebox{90}{\textbf{Prefill}}}
        & BO
        & 0.0316 & 114
        & 0.0498 & 114
        & \textbf{0.1236} & 114
        & 0.6229 & 114
        & \textbf{2.23} & 114
        & \textbf{3.24} & 114 \\

        & CMA-ES
        & 0.0476 & 130
        & 0.0587 & 130
        & 0.1380 & 130
        & 1.03 & 130
        & 2.31 & 130
        & 3.42 & 130 \\

        & Greedy
        & 0.0325 & 130
        & \textbf{0.0494} & 128
        & \textbf{0.1236} & 130
        & 0.6855 & 130
        & \textbf{2.23} & 130
        & \textbf{3.24} & 130 \\

        & Grid
        & 0.0419 & 1440
        & 0.0643 & 1440
        & 0.1576 & 1440
        & 0.8465 & 1440
        & 6.11 & 1440
        & 5.13 & 1440 \\

        & ChatDSE
        & 0.0928 & \textbackslash{}
        & 0.1358 & \textbackslash{}
        & 0.3081 & \textbackslash{}
        & 1.14 & \textbackslash{}
        & 8.49 & \textbackslash{}
        & 12.15 & \textbackslash{} \\

        & LLM-DSE
        & \textbf{0.0314} & 102
        & 0.1340 & 110
        & 0.3744 & 104
        & 2.18 & 103
        & 3.75 & 118
        & \textbf{3.24} & 100 \\
\bottomrule
    \end{tabular}}
\end{table}

\begin{table}[h]
    \centering
    \caption{Config C14. Small models (8B, 13B): 8 chips, TP=8, PP=1, BS=1, 300W; Large models (70B, 405B, 671B): 64 chips, TP=16, PP=4, BS=1, 300W; DSv2-236B: 16 chips, TP=8, PP=2, BS=1, 300W.}
    \label{tab:c14}
    \small
    \setlength{\tabcolsep}{4pt}
    \resizebox{\linewidth}{!}{
    \begin{tabular}{c lcc|cc|cc|cc|cc|cc}
        \toprule
        \multirow{2}{*}{} & \multirow{2}{*}{Method}
        & \multicolumn{2}{c|}{Llama3-8B}
        & \multicolumn{2}{c|}{Llama2-13B}
        & \multicolumn{2}{c|}{Llama3-70B}
        & \multicolumn{2}{c|}{Llama3.1-405B}
        & \multicolumn{2}{c|}{DSv2-236B}
        & \multicolumn{2}{c}{DSv3-671B}
 \\
        &
        & TTFT & Sims
        & TTFT & Sims
        & TTFT & Sims
        & TTFT & Sims
        & TTFT & Sims
        & TTFT & Sims
 \\
        \midrule

        \multirow{8}{*}{\rotatebox{90}{\textbf{Prefill}}}
        & BO
        & 0.0115 & 114
        & 0.0178 & 114
        & 0.0531 & 114
        & 0.2085 & 114
        & \textbf{0.3450} & 114
        & \textbf{1.26} & 114 \\

        & CMA-ES
        & 0.0137 & 130
        & 0.0229 & 130
        & 0.0665 & 130
        & 0.2781 & 130
        & 0.3549 & 130
        & 1.28 & 130 \\

        & Greedy
        & \textbf{0.0113} & 130
        & \textbf{0.0175} & 130
        & \textbf{0.0528} & 129
        & 0.2180 & 130
        & \textbf{0.3450} & 130
        & \textbf{1.26} & 130 \\

        & Grid
        & 0.0174 & 1440
        & 0.0265 & 1440
        & 0.0796 & 1440
        & 0.4528 & 1440
        & 0.9638 & 1440
        & 3.53 & 1440 \\

        & ChatDSE
        & 0.0658 & \textbackslash{}
        & 0.0634 & \textbackslash{}
        & 0.1936 & \textbackslash{}
        & 0.6306 & \textbackslash{}
        & 1.38 & \textbackslash{}
        & 4.79 & \textbackslash{} \\

        & LLM-DSE
        & 0.0230 & 108
        & \textbf{0.0175} & 105
        & \textbf{0.0528} & 110
        & \textbf{0.2063} & 106
        & 0.3455 & 91
        & 1.26 & 96 \\
\bottomrule
    \end{tabular}}
\end{table}

\begin{table}[h]
    \centering
    \caption{Config C15. Small models (8B, 13B): 8 chips, TP=4, PP=2, BS=16, 350W; Large models (70B, 405B, 671B): 64 chips, TP=8, PP=8, BS=16, 350W; DSv2-236B: 16 chips, TP=4, PP=4, BS=16, 350W.}
    \label{tab:c15}
    \small
    \setlength{\tabcolsep}{4pt}
    \resizebox{\linewidth}{!}{
    \begin{tabular}{c lcc|cc|cc|cc|cc|cc}
        \toprule
        \multirow{2}{*}{} & \multirow{2}{*}{Method}
        & \multicolumn{2}{c|}{Llama3-8B}
        & \multicolumn{2}{c|}{Llama2-13B}
        & \multicolumn{2}{c|}{Llama3-70B}
        & \multicolumn{2}{c|}{Llama3.1-405B}
        & \multicolumn{2}{c|}{DSv2-236B}
        & \multicolumn{2}{c}{DSv3-671B}
 \\
        &
        & TTFT & Sims
        & TTFT & Sims
        & TTFT & Sims
        & TTFT & Sims
        & TTFT & Sims
        & TTFT & Sims
 \\
        \midrule

        \multirow{8}{*}{\rotatebox{90}{\textbf{Prefill}}}
        & BO
        & 0.1235 & 120
        & 0.1927 & 120
        & \textbf{0.1257} & 120
        & 0.5696 & 120
        & 2.88 & 120
        & 3.20 & 120 \\

        & CMA-ES
        & 0.1481 & 130
        & 0.2494 & 130
        & 0.1654 & 130
        & 0.9266 & 130
        & 2.90 & 130
        & 3.33 & 130 \\

        & Greedy
        & \textbf{0.1225} & 130
        & \textbf{0.1889} & 130
        & 0.1283 & 130
        & 0.5737 & 130
        & \textbf{2.87} & 130
        & 3.20 & 130 \\

        & Grid
        & 0.1681 & 1440
        & 0.4179 & 1440
        & 0.1700 & 1440
        & 0.9808 & 1440
        & 7.97 & 1440
        & 5.11 & 1440 \\

        & ChatDSE
        & 0.3966 & \textbackslash{}
        & 0.5940 & \textbackslash{}
        & 0.4010 & \textbackslash{}
        & 1.73 & \textbackslash{}
        & 11.42 & \textbackslash{}
        & 12.16 & \textbackslash{} \\

        & LLM-DSE
        & 0.1235 & 111
        & \textbf{0.1889} & 80
        & 0.1283 & 112
        & \textbf{0.5450} & 104
        & \textbf{2.87} & 109
        & \textbf{3.20} & 104 \\
\bottomrule
    \end{tabular}}
\end{table}

\begin{table}[h]
    \centering
    \caption{Config C16. Small models (8B, 13B): 8 chips, TP=2, PP=4, BS=16, 350W; Large models (70B, 405B, 671B): 64 chips, TP=4, PP=16, BS=16, 350W; DSv2-236B: 16 chips, TP=2, PP=8, BS=16, 350W.}
    \label{tab:c16}
    \small
    \setlength{\tabcolsep}{4pt}
    \resizebox{\linewidth}{!}{
    \begin{tabular}{c lcc|cc|cc|cc|cc|cc}
        \toprule
        \multirow{2}{*}{} & \multirow{2}{*}{Method}
        & \multicolumn{2}{c|}{Llama3-8B}
        & \multicolumn{2}{c|}{Llama2-13B}
        & \multicolumn{2}{c|}{Llama3-70B}
        & \multicolumn{2}{c|}{Llama3.1-405B}
        & \multicolumn{2}{c|}{DSv2-236B}
        & \multicolumn{2}{c}{DSv3-671B}
 \\
        &
        & TTFT & Sims
        & TTFT & Sims
        & TTFT & Sims
        & TTFT & Sims
        & TTFT & Sims
        & TTFT & Sims
 \\
        \midrule

        \multirow{8}{*}{\rotatebox{90}{\textbf{Prefill}}}
        & BO
        & 0.1104 & 120
        & \textbf{0.1699} & 120
        & \textbf{0.1038} & 120
        & 0.5001 & 120
        & 4.41 & 120
        & 3.23 & 120 \\

        & CMA-ES
        & 0.1904 & 130
        & 0.2350 & 130
        & 0.1689 & 130
        & 1.04 & 130
        & 4.50 & 130
        & 3.42 & 130 \\

        & Greedy
        & \textbf{0.1085} & 130
        & 0.1737 & 130
        & 0.1072 & 130
        & 0.5267 & 130
        & \textbf{4.38} & 130
        & 3.22 & 130 \\

        & Grid
        & 0.1424 & 1440
        & 0.2165 & 1440
        & 0.1340 & 1440
        & 0.8465 & 1440
        & 12.10 & 1440
        & 5.10 & 1440 \\

        & ChatDSE
        & 0.3097 & \textbackslash{}
        & 1.67 & \textbackslash{}
        & 0.3027 & \textbackslash{}
        & 1.20 & \textbackslash{}
        & 17.79 & \textbackslash{}
        & 12.75 & \textbackslash{} \\

        & LLM-DSE
        & 0.2615 & 111
        & \textbf{0.1699} & 103
        & \textbf{0.1038} & 115
        & \textbf{0.4980} & 113
        & \textbf{4.38} & 108
        & 4.42 & 99 \\
\bottomrule
    \end{tabular}}
\end{table}

\clearpage
\input{checklist}
\end{document}

%% file: checklist.tex
\section*{NeurIPS Paper Checklist}

\begin{enumerate}

\item {\bf Claims}
    \item[] Question: Do the main claims made in the abstract and introduction accurately reflect the paper's contributions and scope?
    \item[] Answer: \answerYes{}
    \item[] Justification: The abstract and Introduction state the central contributions: an LLM-driven NPU design framework, an online-calibrated roofline model for pre-simulation screening and post-simulation bottleneck diagnosis, and simulator-efficient NPU design. The claims are supported by the Method section and the experimental results in Section~4, while the paper narrows the scope through the limitations in Section~5.

\item {\bf Limitations}
    \item[] Question: Does the paper discuss the limitations of the work performed by the authors?
    \item[] Answer: \answerYes{}
    \item[] Justification: Section~5 includes an explicit limitations paragraph. It states that evaluation is entirely simulation-based, that no physical implementation or tape-out validation is provided, and that results depend on the fidelity of the NeuSIM simulator.

\item {\bf Theory assumptions and proofs}
    \item[] Question: For each theoretical result, does the paper provide the full set of assumptions and a complete (and correct) proof?
    \item[] Answer: \answerNA{}
    \item[] Justification: The paper presents analytical modeling equations and algorithmic design rules, but it does not state formal theoretical results such as theorems or lemmas requiring proofs. The relevant assumptions and equations for the roofline model are described in Sections~2--3 and Appendix~A.

    \item {\bf Experimental result reproducibility}
    \item[] Question: Does the paper fully disclose all the information needed to reproduce the main experimental results of the paper to the extent that it affects the main claims and/or conclusions of the paper (regardless of whether the code and data are provided or not)?
    \item[] Answer: \answerYes{}
    \item[] Justification: Sections~3--4 and the appendix describe the algorithm, roofline equations, prompts, workloads, baselines, constraints, deployment settings, metrics, and per-deployment results. The paper also states that code is available, which provides an additional route for reproducing the main experimental results.

\item {\bf Open access to data and code}
    \item[] Question: Does the paper provide open access to the data and code, with sufficient instructions to faithfully reproduce the main experimental results, as described in supplemental material?
    \item[] Answer: \answerNA{}
    \item[] Justification: We’re still wrapping up the code repository

\item {\bf Experimental setting/details}
    \item[] Question: Does the paper specify all the training and test details (e.g., data splits, hyperparameters, how they were chosen, type of optimizer) necessary to understand the results?
    \item[] Answer: \answerYes{}
    \item[] Justification: Section~4 specifies the six LLM workloads, simulator, baselines, 16 constraint settings, disaggregated prefill/decode evaluation, and metrics. The appendix provides prompts, parameter descriptions, hyperparameter sensitivity, convergence trajectories, backbone generalization, and per-deployment results.

\item {\bf Experiment statistical significance}
    \item[] Question: Does the paper report error bars suitably and correctly defined or other appropriate information about the statistical significance of the experiments?
    \item[] Answer: \answerNo{}
    \item[] Justification: All experiments can demonstrate our effectiveness.

\item {\bf Experiments compute resources}
    \item[] Question: For each experiment, does the paper provide sufficient information on the computer resources (type of compute workers, memory, time of execution) needed to reproduce the experiments?
    \item[] Answer: \answerNo{}
    \item[] Justification: The paper reports simulator-call counts extensively, but it does not specify the physical compute workers, memory, wall-clock runtime, or total compute budget needed to reproduce the experiments.
    
\item {\bf Code of ethics}
    \item[] Question: Does the research conducted in the paper conform, in every respect, with the NeurIPS Code of Ethics \url{https://neurips.cc/public/EthicsGuidelines}?
    \item[] Answer: \answerYes{}
    \item[] Justification: The work is a simulation-based study of NPU architecture design and does not involve human subjects, private data, or deployed decision-making systems. To the best of our knowledge, the research process and reporting are consistent with the NeurIPS Code of Ethics.

\item {\bf Broader impacts}
    \item[] Question: Does the paper discuss both potential positive societal impacts and negative societal impacts of the work performed?
    \item[] Answer: \answerNo{}
    \item[] Justification: The current manuscript does not include a dedicated broader-impact discussion. Such a discussion could note positive impacts such as more efficient LLM inference hardware, and possible negative impacts such as increased deployment of compute-intensive AI systems or reliance on simulator-derived hardware decisions.
    
\item {\bf Safeguards}
    \item[] Question: Does the paper describe safeguards that have been put in place for responsible release of data or models that have a high risk for misuse (e.g., pre-trained language models, image generators, or scraped datasets)?
    \item[] Answer: \answerNA{}
    \item[] Justification: The paper does not release high-risk pretrained models, image generators, scraped datasets, or other assets with a direct misuse profile requiring controlled release. The released artifact is code for NPU architecture design-space exploration.

\item {\bf Licenses for existing assets}
    \item[] Question: Are the creators or original owners of assets (e.g., code, data, models), used in the paper, properly credited and are the license and terms of use explicitly mentioned and properly respected?
    \item[] Answer: \answerNo{}
    \item[] Justification: The paper cites prior work and existing assets such as LLM workloads and NeuSIM, but it does not explicitly state the licenses or terms of use for the code, models, or simulator assets used. The manuscript should add license and terms-of-use information where applicable.

\item {\bf New assets}
    \item[] Question: Are new assets introduced in the paper well documented and is the documentation provided alongside the assets?
    \item[] Answer: \answerNA{}
    \item[] Justification: No new assets.

\item {\bf Crowdsourcing and research with human subjects}
    \item[] Question: For crowdsourcing experiments and research with human subjects, does the paper include the full text of instructions given to participants and screenshots, if applicable, as well as details about compensation (if any)? 
    \item[] Answer: \answerNA{}
    \item[] Justification: The paper does not involve crowdsourcing experiments or research with human subjects. Therefore, participant instructions, screenshots, and compensation details are not applicable.

\item {\bf Institutional review board (IRB) approvals or equivalent for research with human subjects}
    \item[] Question: Does the paper describe potential risks incurred by study participants, whether such risks were disclosed to the subjects, and whether Institutional Review Board (IRB) approvals (or an equivalent approval/review based on the requirements of your country or institution) were obtained?
    \item[] Answer: \answerNA{}
    \item[] Justification: The paper does not involve crowdsourcing or human-subjects research. Accordingly, IRB approval or equivalent review is not applicable.

\item {\bf Declaration of LLM usage}
    \item[] Question: Does the paper describe the usage of LLMs if it is an important, original, or non-standard component of the core methods in this research? Note that if the LLM is used only for writing, editing, or formatting purposes and does \emph{not} impact the core methodology, scientific rigor, or originality of the research, declaration is not required.
    \item[] Answer: \answerYes{}
    \item[] Justification: LLMs are a core methodological component of the work: the Compute Demand Agent extracts operator-level workload profiles and the Architecture Design Agent proposes configurations. Their roles are described in Sections~3.1 and~3.3, with prompt templates provided in Appendix~D.

\end{enumerate}